\documentstyle[preprint,prd,aps,floats,epsf]{revtex}

\newcommand{\postscript}[2]
{\setlength{\epsfxsize}{#2\hsize}
\centerline{\epsfbox{#1}}}

\tighten

\begin{document}

\draft

\preprint{
\noindent
\begin{minipage}[t]{3in}
\begin{flushleft}
May 1996
\end{flushleft}
\end{minipage}
\hfill
\begin{minipage}[t]{3in}
\begin{flushright}
TUM--HEP--245/96\\
SFB--375/97\\
LMU--TPW--96--09
\end{flushright}
\end{minipage}
}


\title{Supersymmetric Neutrino Masses, R Symmetries,\\
 and The Generalized $\mu$ Problem}

\author{Hans-Peter Nilles}
\address{\it Physics Department, Technische 
Universit$\ddot{a}$t M$\ddot{u}$nchen \\
D-85748 Garching, Germany\\
and\\
Werner-Heisenberg-Institut\\
Max-Planck-Institut f$\ddot{u}$r Physik, Werner-Heisenberg-Institut\\
D-80805 M$\ddot{u}$nchen, Germany}
\author{Nir Polonsky}
\address{\it
Sektion Physik der
Universit$\ddot{a}$t M$\ddot{u}$nchen \\
Theoretische Physik-Lehrstuhl Prof.\ Wess, Theresienstrasse 37\\
D-80333 M$\ddot{u}$nchen,  Germany}


\maketitle

\begin{abstract}
In supersymmetric models a tree-level neutrino mass could originate
from the (weak-scale) superpotential.
We propose and examine a realization of that idea,
which arises naturally in the framework of a spontaneously
broken $U(1)$ $R$-symmetry. 
The solution to the neutrino mass problem 
could shed light in this framework on 
the possible resolution of the $\mu$ problem.
Furthermore, the suppression of the neutrino mass
in comparison to the weak scale arises dynamically and  
need not be encoded in the superpotential.
The latter mechanism operates, for example, in universal
models for the soft supersymmetry breaking terms.
Phenomenological and cosmological implications of the model
are also discussed, some of which are shown to hold
more generally.
We also note that future signatures 
could include observable enhancement of dijet and multijet
production rates and a correlation
between the supersymmetric and neutrino spectra.
\end{abstract}


\section{Introduction}
\label{sec:s1}
Within the framework of the Standard Model of
the electroweak and strong interactions 
the neutrinos are massless to all orders in perturbation theory.
However, it is widely believed, based
on the interpretation of current observations, that
the neutrinos are massive and light,
i.e., $m_{\nu} \sim {\cal{O}}($1 - 100 eV) \cite{langrev}.
If confirmed by future
experiments (e.g., the next generation of underground observatories and 
the long-baseline oscillation experiments)
the massive neutrinos would provide an unambiguous signal of
physics beyond the Standard Model (SM).
For example, the small neutrino masses are often
attributed to some sort of  a see-saw mechanism involving
an intermediate scale  
$M_{\mbox{\tiny Intermediate}} \sim {\cal{O}}(10^{12}$ GeV)
right-handed neutrino (or some other structure at that scale),
i.e., $m_{\nu} \approx M_{\mbox{\tiny weak}}^{2}/
M_{\mbox{\tiny Intermediate}}$
 \cite{seesaw}. The neutrino mass in that case is a signal
of the physics at intermediate scales.

A different and most interesting possibility is that 
the neutrino mass is  a signal of weak-scale physics.
Indeed, this could be the case within a supersymmetric framework
\cite{nillesrev,hk,books} as it could incorporate
a new and novel 
(but often overlooked) mechanism for the generation
of neutrino masses which was first proposed by Hall and Suzuki \cite{hs}.
Supersymmetric theories  
naturally accommodate weak-scale  scalars 
and are a strongly motivated  candidate for an extension
of the SM at a scale $M_{SUSY} \sim {\cal{O}}(100$ GeV$- 1$ TeV) 
\cite{nillesrev,hk,books}. 
Thus, an investigation of neutrino mass generation within this
framework is well motivated
and would provide an alternative and unique
interpretation of a neutrino mass signal.
In turn, the latter could provide valuable information
on the supersymmetric structure both at weak and Planckian scales.
In this work\footnote{A preliminary version of this work
was presented by N. P. 
in the {\it Summer Institute on Signals of Unified Theories},
Laboratori Nazionali del Gran Sasso, Italy, September, 1995.} 
we will propose a new realization of that mechanism,
argue for its naturalness within
the context of supersymmetric model building,
and examine its phenomenological
and cosmological  implications.
(We will assume the supersymmetric extension
of the SM throughout this work.)

The neutrino mass generation 
follows in a straightforward manner if one allows
a low-energy (bilinear) mass 
superpotential, $W_{M}$, 
which is the most general one in the 
fields of the minimal supersymmetric extension (MSSM)
and the SM 
$SU(3)_{c}\times SU(2)_{L}\times U(1)_{Y}$
gauge group. The (renormalizable) superpotential reads in that case
\begin{equation}
W = \mu_{\alpha}L^{\alpha}H_{2} + W_{Y},
\label{sp1}
\end{equation}
where\footnote{Hereafter we denote (suppressing family indices)
the $SU(2)_{L}$ quark and lepton doublet
chiral superfields by $Q$ and $L$, the up- and down-quark singlets
by $U$ and $D$, and the lepton singlets by $E$, respectively.
$H_{1}$  and $H_{2}$ 
denote the $Y = -\frac{1}{2}$ and $+\frac{1}{2}$ Higgs doublets, respectively.
A chiral superfield
and its scalar component 
will be denoted below by the same symbol.} 
$L_{\alpha} = (H_{1},L_{\tau},L_{\mu},L_{e})$ 
is [listing $SU(3)_{c}\times SU(2)_{L}\times U(1)_{Y}$ quantum numbers]
the (1,2,-1/2) four-vector in field space, the four-vector
$\mu_{\alpha}$ is of the order of magnitude of the weak scale and $M_{SUSY}$,
and $W_{Y}$
is the (trilinear) Yukawa part of the superpotential. 
One neutrino species (see below) is now massive at tree level
as a result of the superpotential mass term 
(in a similar manner to the fermion partners
of the Higgs bosons -- the Higgsinos). 
The details of the Yukawa superpotential $W_{Y}$ would only
affect the loop-level corrections to the masses.

It is quite striking that the SM massless fermions -- the neutrinos --
are the only SM fermions that
are allowed by the symmetries to have a supersymmetric mass.
In general,
supersymmetrtic mass parameters are naturally
of the order of the grand-scale or are zero,
e.g., in string theory and grand-unified models.
Hence, low-energy mass parameters 
(which are of the order of the weak-scale) in (\ref{sp1}) 
represent a perturbation to the above expectation and
they parameterize the high-energy physics
in a similar way to the dimensionful supersymmetry breaking
soft parametrs which set the scale $M_{SUSY}$
(see below for a discussion of the $\mu$-problem).
Once mass parameters are introduced, 
the term $\mu_{\alpha}L^{\alpha}H_{2}$ (rather than
only $\mu_{H_{1}}H_{1}H_{2}$, which  is the only superpotential
mass term included in the MSSM case) is the most natural 
choice. The neutrino mass then arises 
from  weak-scale parameters, but in fact, parameterizes 
the high-energy physics.
Naively, one might have  expected 
that (unless $\mu_{L_{\tau,\, \mu,\, e}} \rightarrow 0$) 
$m_{\nu} \approx {\cal{O}}(|\mu_{\alpha}|) \approx {\cal{O}}(M_{SUSY})$. 
However, a simple condition \cite{hs,bn}, which, 
as we will show, 
may be realized dynamically in universal models
for the soft supersymmetry breaking parameters,
guarantees that this is not the case
and that neutrinos as light as 
${\cal{O}}($100 eV$)$ are obtained.

Other proposals for weak-scale neutrino mass generation
in a supersymmetric framework
\cite{lee,aulakh,eg,rv,yukawaloop1,yukawaloop2} include
mass generation due to the soft supersymmetry breaking sector,
spontaneous lepton number breaking, and Yukawa interactions.
The supersymmetric neutrino mass mechanism \cite{hs,dawson}, 
that we will examine here, 
embodies various aspects of all other proposals.

The neutrino mass generation involves, 
in our case, several delicate issues which require
some elaboration. While doing so, we will establish
the outline of our proposed  models.

{\it R-parity violation:}
The superpotential $W$ given in (\ref{sp1}) 
does not preserve lepton number, and thus, the  discrete 
$Z_{2}$ $R$-parity \cite{z2} $R_{P} = (-)^{2S + 3B + L}$ 
(where $S$, $B$, and $L$ are the particle spin, baryon and lepton
numbers, respectively) that is typically
imposed in the MSSM 
(and forbids weak-scale neutrino mass generation)
is also broken.
Unlike in generic broken $R$-parity models, the breaking
here is restricted and does not lead  to unacceptable 
proton-decay rates. Specifically, 
the  (tree-level) neutrino mass generation
is insensitive to the details of $W_{Y}$,
and the Yukawa superpotential
could  still have (approximately) 
its SM form and 
preserve this or some other symmetry
 so that the proton is long lived. 
(In fact, for that purpose it is sufficient
to preserve  at the renormalizable level only baryon number,
e.g., this is the case
in the $Z_{3}$ baryon parity \cite{z3} and similar \cite{kmn} models.)
For our purpose it is enough to consider 
$W_{Y} = W_{Y}^{\mbox{\tiny MSSM}}$ where
(suppressing family indices)
\begin{equation}
W_{Y}^{\mbox{\tiny MSSM}} = 
h_{U}H_{2}QU +  h_{D}H_{1}QD + h_{E}H_{1}LE.
\label{wy}
\end{equation} 
Below, we will show that this is indeed a natural 
choice in the models that we consider.

{\it Rotations and lepton-number redefinitions:}
Lepton number violation
can be rotated from $W_{M}$ onto $W_{Y}$, introducing
the lepton number violating terms\footnote{
In the case of an arbitrary
$W_{Y}^{\mbox{\tiny LNV}}$ 
the Yukawa couplings are sometimes
denoted by $\lambda =  h_{E}^{\mbox{\tiny LNV}}$
and $\lambda ' =  h_{D}^{\mbox{\tiny LNV}}$.
Our notation aims at stressing the relation in our case 
between $W_{Y}^{\mbox{\tiny LNV}}$  and $W_{Y}^{\mbox{\tiny MSSM}}$ 
(see below).}
\begin{equation}
W_{Y}^{\mbox{\tiny LNV}} = 
 h_{D}^{\mbox{\tiny LNV}}LQD + \frac{1}{2}h_{E}^{\mbox{\tiny LNV}}
[L, L]E,
\label{wlnv}
\end{equation} 
where commutation with respect to generation
indices is implied in the second Yukawa term (which  vanishes
in the  case of one lepton generation). 
$R$-parity is now replaced, e.g., by baryon parity.
Such rotations are useful when discussing the 
implications of the model
at low energies (where all relevant symmetries are broken).
The new Yukawa couplings are not arbitrary in that case and 
lead to the same phenomenology.
However, rotations are impossible if $W$
is invariant under a symmetry which does not commute
with the $SU(4)$ symmetry of $L_{\alpha}$ rotations in the field space,
i.e., a symmetry which  distinguishes the Higgs and
lepton superfields. 
We will assume that this is the case
at  high energies, e.g., at Planckian scales. 
We will show later that  such a  ``non-commuting'' symmetry 
is, in fact, a desired feature and that it can be
related to the ad hoc absence  of an arbitrary 
$W_{Y}^{\mbox{\tiny LNV}}$ (and also to the absence of
lepton number violation in the soft terms).
In particular, one may not need to impose an additional
symmetry on $W_{Y}$.
Note also that the rotation leading to (\ref{wlnv}) is not scale invariant
and that $W_{Y}^{\mbox{\tiny LNV}} \neq 0$
will regenerate  lepton number violation in $W_{M}$ via renormalization
group scaling. It suggests that one can define the SM lepton number
consistently only at low-energies.
For that reason,  and as a result of our symmetry assumption,
we distinguish
between the high- (MSSM) and low-energy (SM) definitions of the leptons. 
The former are defined by the superpotential (\ref{wy}) while  
the latter will be  chosen 
(after all symmetries are broken
and rotations are possible) in most cases
as the  perpendicular directions to
the relevant weak-scale expectation value.
This distinction will prove as a useful model-building tool.

{\it The $\mu$-problem and the choice of a symmetry:}
An arbitrary (high-energy) vector with  a magnitude 
$|\mu| \sim {\cal{O}}(M_{SUSY})$
could lead to highly suppressed neutrino masses and
there is no need to significantly suppress 
$\mu_{L_{\tau}}/\mu_{H_{1}}$ etc. (this will be demonstrated 
in the following sections).
In particular, the neutrino mass is related in our framework to the solution
of the celebrated $\mu$-problem \cite{muproblem}
to which we alluded above, i.e., explaining
$|\mu|/M_{Planck} \rightarrow 0$ and 
$|\mu| \sim M_{SUSY} > 0$, which is the manifestation of
the correlation between the supersymmetry preserving 
[with a natural scale  of ${\cal{O}}(M_{Planck})$]
and softly breaking [with an ad hoc scale  
$M_{SUSY} \sim {\cal{O}}(M_{Z})$)] sectors.
The correlation is imposed by requiring 
the correct electroweak symmetry breaking pattern \cite{alex}.
(Usually the problem is phrased in terms of $|\mu_{H_{1}}|$.)
We will adopt a somewhat  ambitious approach and 
require  that 
the solution to the $\mu$-problem 
is determined by the same symmetry that was proposed above
[and which does not commute with the $SU(4)$].
In particular, we will consider a continuous $U(1)_{R}$ R-symmetry 
(the $R$-charge is defined as above)
that is known to be relevant for the solution of the 
$\mu$-problem \cite{gm,nk,muzero}, and
under which $\mu_{H_{1}}$ and $\mu_{L_{\tau,\,\mu,\,e}}$
effectively carry different $R$-charges. 
In principle, we could use other $U(1)$ symmetries to break the $SU(4)$,
e.g., a Peccei-Quinn symmetry \cite{pq}
or the stringy inspired $U(1)$'s of Ref.\ \cite{cvlang}.
In the latter case $\mu_{\alpha}$ could be forbidden by the new $U(1)$
gauge symmetry but  induced by weak-scale expectation values 
in a similar manner to the authors' original proposal for the generation
of $\mu_{H_{1}}$. We will concentrate in this work on the
$U(1)_{R}$ case.
Generalization of our proposal to models in which other symmetries
play a similar role in solving the $\mu$-problem is straightforward.

{\it Operator classification:}
In the symmetry  framework $\mu_{H_{1}}$  and $\mu_{L_{\tau,\,\mu, \,e}}$
are realized as nonrenormalizable operators (NRO's)
present  in the effective
low-energy superpotential \cite{muproblem}.
The $R$-symmetry allows us to classify three categories 
of mass terms in $W$
[which is normalized\footnote{
We assume the $U(1)_{R}$ transformation law
$\phi \rightarrow e^{i\alpha R}\phi$,
$\psi \rightarrow e^{i\alpha(R - 1)}\psi$,
and $F \rightarrow e^{i\alpha(R - 2)}F$,
for the scalar, fermion and auxiliary components of
a chiral superfield $\Phi$ with R-charge $R$, respectively.
}
 to carry $R-$charge $R(W) = 2$]:
\begin{mathletters}
\label{mus}
\begin{equation}
\mu_{H}\,\,\,\,\,\, R(\mu_{H}) = 2 \,\,\,\,\,\, {\mbox{operator: }} H_{1}H_{2},
\label{mus1}
\end{equation}
\begin{equation}
\mu_{L}\,\,\,\,\,\, R(\mu_{L}) = 1\,\,\,\,\,\, {\mbox{operator: }} 
L_{\tau,\, \mu,\, e}H_{2},
\label{mus2}
\end{equation}
\begin{equation}
\mu_{N} \,\,\,\,\,\, R(\mu_{N}) = 0 \,\,\,\,\,\, {\mbox{operator: }} N^{2};\,\Phi \bar{\Phi},
\label{mus3}
\end{equation}
\end{mathletters}
where we have assumed the standard $R$-charge assignments as above 
(substituting $S = 0$), i.e.,
$R(H_{1,\,2})   = 0$ and 
$R(Q,\, U,\, D,\, L,\, E,\, N,\, \Phi,\, \bar{\Phi}) = 1$.
Thus, the $R$-symmetry provides us with a convenient book-keeping tool.
Note that the proposal $\mu_{H_{1}} \sim \langle W \rangle/M_{P}^{2}$
\cite{cm} is trivially realized in this framework.
Generically (but see exceptions below), one expects
that the higher $R$-charge $\mu$ carries the more suppressed it is.
Furthermore, since lepton number violation is in the form 
of nonrenormalizable operators, 
and based on dimensional arguments, one also expects
that, e.g., $h_{E}^{\mbox{\tiny LNV}} \sim |\mu_{L}|/M_{P} \rightarrow 0$.
[Note that $R(h^{\mbox{\tiny LNV}}) = -1$.]
Thus, $W_{Y}$ is effectively $W_{Y}^{\mbox{\tiny MSSM}}$
with an accidental discrete $R_{P}$ symmetry.
Similar arguments hold for the dimensionless couplings in
the Kahler potential, and typically lepton number violation
in the soft terms is also suppressed. 
As a result of the symmetry selection rules,
$R$-number violation, and in particular, lepton number violation, 
is contained in $W_{M}$ as advocated.  
The singlet $\mu$-parameters may not be suppressed, 
and the right handed neutrino
$N$ (or other singlets) and vector-like exotic $\Phi$ and $\bar{\Phi}$ 
supermultiplets with
standard $R$-charge assignments are expected to be heavy.
(The latter points are, in fact, relevant
for the dynamical suppression of the
neutrino masses discussed below.)

We are now in a position to outline our proposal:
The $U(1)_{R}$ symmetry selection rules
require that there is a  $\sim M_{Z} / M_{Planck}$  hierarchy 
between the symmetry violation
in the supersymmetric mass terms and in the 
Yukawa terms and the Kahler potential.
As a result, it leads to the generation of  $\mu$ terms
of the right magnitude, accidental
$R_{P}$ symmetry in the Yukawa terms,  
and to a consistent definition of the low-energy lepton number.
In addition, the neutrino masses are suppressed dynamically.

We will examine the above issues in greater detail
and expand our discussion in the following sections.
In Section II we briefly discuss 
possible realizations (and their problems) 
of the $U(1)_{R}$ symmetry, and present some simple examples that
realize (or are exceptions to)  
the main features discussed above, i.e., the suppression
of non-minimal Yukawa interactions and the hierarchy among 
the different $\mu$ terms.
We will show that the nature of the hierarchy depends on whether
or not the relevant operators involve $F$-terms.
In Section III  we discuss the neutral fermion mass matrix
and the sufficient conditions for small neutrino masses,
which include the alignment (in field space) of 
$\langle L^{0}_{\alpha} \rangle$ with $\mu_{\alpha}$.
We then proceed and show in Section IV that 
the latter condition and 
$m_{\nu} \ll M_{Z}$ are  achieved in a straightforward manner
in universal models with radiative (electroweak)
symmetry breaking (RSB)  \cite{nillesrev}
via the ``dynamical alignment'' mechanism.
(For simplicity, we discuss a model with only the third generation of leptons
and quarks.) It is also pointed out that grand-unification relations,
as well as an intermediate-scale right-handed neutrino, generically
destabilize this result. 
In section V we  show that lepton number (and individual lepton number)
violations in the model are typically proportional to either the neutrino mass
or to small MSSM Yukawa couplings,
and thus, they are suppressed. We briefly address the issue of possible
discovery channels,
and stress that there could be an 
enhancement of dijet and multijet cross-sections.
Cosmology and astrophysics implications are discussed in section VI.
It is shown that 
weak-scale neutrino mass generation implies that the 
supersymmetric partners (superpartners)
of ordinary matter and gauge fields are not
stable on cosmological time scales.
We will alternate in our discussion between the high-energy 
(Sections II and IV)
and low-energy (Sections III, V and VI) 
definitions of the leptons and lepton number, 
as appropriate.
In Section VII we comment on possible family dependencies
in more complete models and on their implications.

We conclude in section VIII, where we suggest that the
proposal for supersymmetric 
neutrino mass generation is most elegantly understood
in terms of symmetry principles that are also relevant to the 
(generalized) $\mu$-problem, 
can be naturally incorporated in simple supersymmetric models,
removes the need to ad hoc generate 
an intermediate scale for the right-handed neutrinos, e.g.,  in string models,
and requires one to consider unorthodox scenarios for supersymmetric
particle astrophysics, cosmology,  and for superpartner collider signatures.

\section{$R$-symmetries and selection rules}
\label{sec:s2}

We have chosen a spontaneously broken $U(1)_{R}$
symmetry as our  primary tool in deriving selection rules
for the low-energy effective theory.
We also have shown above that it allows one to distinguish
three classes
of mass (i.e., $\mu$-) parameters in the effective low-energy superpotential;
it distinguishes  the (high-scale) lepton and Higgs doublets;
and that, on the basis of dimensional arguments,
non-singlet (i.e., lepton-number violating) Yukawa couplings 
are typically suppressed by an inverse power of the Planck mass
in comparison to non-singlet $\mu$-parameters.
The presence  
of such a symmetry is intimately connected
to (dynamical) supersymmetry breaking  \cite{dynamical,ns}
and the $R$-symmetry  could play a natural rule in the solution of
the (either MSSM or our generalized) $\mu$-problem\footnote{
$\mu_{\alpha} \equiv 0$
if the $U(1)_{R}$ symmetry is unbroken 
and the $\mu$-problem is trivially solved \cite{muzero}.
However, such models are strongly constrained
and face many phenomenological (e.g., generation of a gluino mass)
and model-building (e.g., realization of RSB) difficulties 
(see, for example, Ref.\ \cite{fengetal}).}  
\cite{gm,nk}.
Hence, our choice of ``book-keeping'' tool is well motivated.

However, similarly to the Peccei-Quinn case \cite{pq}, 
a spontaneously broken 
continuous $R$-symmetry implies 
an unwanted \cite{axion} (pseudo) Goldstone boson --
the $R$-axion.
The troublesome presence of the $R$-axion
can be resolved $(i)$
if it is an invisible axion, i.e., the $R$-symmetry is broken
at a scale $\theta \gtrsim 10^{10}$ GeV 
(from stellar evolution) 
and below the Planck scale $\theta \lesssim 10^{12}$ 
GeV (from its contribution to the energy density) \cite{kimrev}
(a  higher bound $\theta \lesssim 10^{16}$ may exist in some cases
\cite{axionscale});
$(ii)$ if  the axion receives its mass from NRO's that explicitly
break the symmetry\footnote{It was also suggested \cite{bpr}
that the explicit breaking is related to a constant term in the superpotential
which carries no $R$-charge 
and which is perhaps needed to cancel the vacuum energy. 
In that case the axion mass is related to the size of the 
soft supersymmetry breaking parameters.
In fact, explicit breaking by a large constant
in the gravitational sector is a common solution to the
$U(1)_{R}$ problem  in supergravity models, e.g.,  
in the Polonyi model \cite{polonyi}.}
\cite{ns};
$(iii)$ if the symmetry is gauged (and anomaly free) 
at Planckian scales \cite{gaugeR}; or $(iv)$ if the axion
is rendered heavy by different means than the Higgs mechanism
which operates in $(iii)$.

The first option could be the natural choice if the low-energy
global symmetry results from an anomalous high-energy
$U(1)$ (e.g., in a string theory), and thus, could be stable with respect
to gravitational corrections.
In that case 
\begin{equation}
10^{10} \lesssim \theta \lesssim 10^{12} - 10^{16} {\mbox{ GeV}},
\label{theta1}
\end{equation}
where the symmetry-breaking scale
$\theta$ 
is the scale parameter that enters NRO's with positive powers,
and thus, controls the size of couplings
in the low-energy effective theory.
If the axion scale is related to a typical supergravity
breaking scale then it naturally falls in this range \cite{gm,nk}.

The second option is motivated by the observation
that global symmetries are not likely
to be exact in the presence of gravity 
or if the $U(1)_{R}$ symmetry is accidental
and due to a higher symmetry and renormalizability
\cite{gravity,bd},
but it implies
that a priory we do not have an handle on the choice of NRO's.
Thus, it may undermine our motivation.
Nevertheless, one could still distinguish
$\cal{O}($MeV$)$ 
operators, that are sufficient to 
generate an acceptable axion mass,
from  ${\cal{O}}(M_{Z})$ operators, which are the relevant ones
in our case.  [Recall that the $\mu_{L}$'s need not
be suppressed necessarily
more than the ordinary $\mu_{H}$ parameter and
could be ${\cal{ O}}(M_{Z})$.]
In this scenario  there are no constraints on $\theta$.
(Note that if the axion is hidden, i.e.,
with only gravitational interactions with ordinary matter,
then effects of any operator 
that may be needed to render it
massive are suppressed in the observable sector
by an additional inverse power of $M_{Planck}$.)

The third option is quite attractive.
The symmetry is anomaly free and 
the Goldstone boson is not an axion. 
However, the anomaly cancelation equations
involve the gauge fermions and
 the gravitino, in addition  to the ordinary, exotic and hidden 
matter fermions (and similarly the $D$-terms).
It was recently shown that the symmetry must be broken at a Planckian
scale so that one can tune to a flat $D$-term direction
$\langle D \rangle = 0$ [and avoid
${\cal{O}}(M_{P})$ masses for ordinary fields] \cite{gaugeR}. Thus,
\begin{equation}
\theta \lesssim M_{P}.
\label{theta2}
\end{equation}
The anomaly equations are difficult to solve and require (many)
new SM singlet fields and/or exotic fields with 
non-trivial SM charge assignments 
(and, in some cases, non-standard and family-dependent
$R$-charge assignments for the ordinary SM fields) \cite{gaugeR}.
While the singlet fields could be hidden in the hidden sector,
${\cal{O}}(M_{P})$ mass terms for the (observable-sector)
exotics are forbidden by
the symmetry, and in the existing examples with exotics there are 
$\cal{O}($TeV$)$ colored fields.
We will not consider explicitly models
with such fields  and will constrain
our investigation to the model with minimal matter content.
However, we note that
unless such fields interact with $L_{\alpha}$, they are irrelevant 
for our purposes and do not alter our discussion.

Lastly,  the fourth option could be realized, e.g., 
if the axion is a hidden-sector  field 
and the $U(1)_{R}$ is anomalous with respect
to a hidden sector QCD group. 
The axion would then acquire
a large mass (of order of the hidden sector
QCD scale $\sim 10^{11}$ GeV, assuming dynamical
supersymmetry breaking in the hidden sector)
\cite{kimrev}. In this case one also expects
\begin{equation}
\theta \sim 10^{11} {\mbox{ GeV}}.
\label{theta3}
\end{equation}

Below, we will not promote any option in particular, but comment
on their different implications where relevant.
We will assume standard $R$-charge assignments, unless
otherwise is specified,
that the scale $\theta$ is in the range
suggested by (\ref{theta1}) -- (\ref{theta3}), 
and that the non-vanishing fields are hidden fields (with only
gravitational interactions with ordinary matter). 
The latter is motivated by the assumption of gauge confinement 
and dynamical supersymmetry breaking in the
hidden sector, and  as discussed  above,
could ease some of the problems\footnote{However,
the hidden axion could lead to cosmological problems
typical to weakly (i.e., gravitaionally) interacting fields
with an intermediate-scale energy density in the potential
(the so-called ``Polonyi problem'') \cite{pp,nkb}.}.

Having discussed the possible frameworks in which 
the symmetry can be realized, we now turn to a discussion of some
examples. NRO's can be induced
in our case by scalar vacuum expectation values
(vev's), non-vanishing $F$-terms,
and fermion condensates \cite{nk}, leading to many possible
scenarios which would relate differently
to supersymmetry breaking.  
Since we assume that only SM hidden singlets
participate, then the 
operators (which are nonrenormalizable in that case)
are suppressed by powers
of the Planck mass.
For simplicity, we will assume that 
$M \approx M_{Planck}/\sqrt{8\pi} \approx {\cal{O}}(10^{18}$ GeV) 
is the only large scale suppressing the operators.
Note also that while the form of the operators (i.e., the selection rules)
is dictated by the symmetries [and in particular, $U(1)_{R}$],
their non-vanishing values (i.e., the vev's) may be 
a result of $U(1)_{R}$  breaking or of the breaking of a different symmetry
at a lower scale.
In the latter case 
$\theta$ could be lower then the $U(1)_{R}$ scale. 
[This observation is most relevant if the gauged 
$U(1)_{R}$ option is realized.]
 
\subsection{Scalar vev's and fractional $R$-charges}

The most simple example is the case in which 
$\theta_{i} = \langle z_{i} \rangle$
is given by the vacuum expectation value of the scalar component
$z_{i}$ of a hidden sector chiral superfield $Z_{i}$ with $R$-charge $R_{i}$. 
The $\mu$-parameters, $\mu_{I} = \mu_{H,\, L,\, N}$ for $I=1,\, 2,\, 3$,
respectively, are given in that case by
(omitting hereafter dimensionless couplings and coefficients
in the nonrenormalizable operators)
\begin{equation}
\mu_{I}  = \frac{\prod_{i_{I}= 1}^{N_{I}} \langle z_{i_{I}} \rangle^{n_{i_{I}}}}
{M^{[(\sum_{i_{I} = 1}^{N_{I}} n_{i_{I}}) - 1]}}.
\label{e1}
\end{equation}

We first discuss the most simple case in which
each of the $\mu_{I}$'s depends only on one field, i.e.,  
$R_{1_{I}} = 2/n_{1_{1}},\, 1/n_{1_{2}},\, 0$ for $z_{1_{1},\, 1_{2},\, 1_{3}}$
($\mu_{H,\, L,\, N}$), respectively. 
$\mu_{N}$  is also suppressed by negative powers of $M$ because
$Z$  is a hidden sector field (otherwise, all symmetries
allow a renormalizable term $Z_{1_{3}}NN$ in the supepotential).
However, it need not be  suppressed by more than one power of $M$,
i.e., $\mu_{N} \sim \langle z_{1_{3}}\rangle^{2}/M$. It could be a weak-scale 
or a Planck-scale parameter, depending if 
$\langle z_{1_{3}}\rangle \sim 10^{10}$ GeV
or $\sim M$. If there are no $R = 0$ fields then $\mu_{N} = 0$.
$\mu_{H}$ and $\mu_{L}$ are weak-scale parameters, leading to the constraint
$n_{1_{1},\, 1_{2}} \sim [\ln(M_{Z}/M)]/[\ln(\langle z_{1_{1},\,1_{2}}\rangle/M)]$,
i.e., $n = 2$ for  $\langle z \rangle \sim 10^{10}$ GeV and
$n = 8$ for  $\langle z \rangle \sim 10^{16}$ GeV.
If $\mu_{H}$ and $\mu_{L}$ are both given by the same field $z_{1}$,
then $n_{1_{1}}$ must be even and 
$\mu_{L} = \sqrt{M_{P}\mu_{H}}$. 
In that case the electroweak symmetry
breaking is induced by the scalar neutrinos (sneutrinos)  
rather than $H_{1}$ which is now decoupled from $H_{2}$. 
(We discuss electroweak symmetry breaking in section IV.)

More generally, the operators could involve several fields
such that $\sum_{i_{I} = 1}^{N_{I}} n_{i_{I}}R_{i_{I}} = 2,\, 1,\, 0$, 
for $I=1,\, 2,\, 3$, respectively.
Let us first assume, however,  that all (non-vanishing) singlets have positive
$R$ charges. Then (because of the holomorphicity of the superpotential)
the situation is similar to the one field case, i.e., 
$\mu_{H}  < {\mu_{L}}$ (and $\mu_{N} = 0$ unless there is 
a non-vanishing $R = 0$ field). More importantly, $h^{\mbox{\tiny LNV}} = 0$
in this case.

Lastly, there could be fields with negative $R$-charges
[in particular, if the $U(1)_{R}$ is gauged and anomaly free]. 
This most general case 
could allow for, e.g., $\mu_{L} = \mu_{H}\times[\langle z \rangle/M]^{l}
< \mu_{H}$ and $ R (Z)= -1/l$
[and also $\mu_{N} = \langle z \rangle \langle z' \rangle /M$
with $R(Z') = -R(Z)$].
However, one could also have now 
$h^{\mbox{\tiny LNV}} \sim h^{\mbox{\tiny BNV}} \sim \mu_{L}/\mu_{H}$
($h^{\mbox{\tiny BNV}}$ is a baryon number violating coupling).
Both, $\mu_{L}$ and $h^{\mbox{\tiny LNV}}$  
(as well as $h^{\mbox{\tiny BNV}}$)  
vanish in the MSSM limit $l \gg 1$.
Otherwise, one has to impose an additional symmetry on
$W_{Y}$. For example, $\mu_{H}$, $\mu_{L}$ and $[\langle z \rangle/M]$
have  (standard) Peccei-Quinn charges\footnote{We assume a Peccei-Quinn
charge $PQ = 1$ for $H_{1}$ and $H_{2}$ and $PQ = -1/2$
for all other ordinary matter.
The anomalous $U(1)$ would be given in that case
by some linear combination of the (anomalous) 
$U(1)_{PQ}$ and $U(1)_{R}$.}
 of $-2,\, -1/2$ and $3/2l$, respectively.
Thus, a Peccei-Quinn symmetry would forbid in this case
the dangerous Yukawa couplings.

\subsection{A possible hierarchy between $\mu_{\alpha}$ and 
$\partial\mu_{\alpha}$}
The chiral superfields discussed in the previous example
could also have non-vanishing $F$-terms which are of the
order of magnitude of the 
supersymmetry breaking scale, $F_{Z} = {\cal{O}}(M_{SUSY}M)$ 
(i.e., they contribute to supersymmetry breaking).
In that case, $Z$ could also provide a source for the soft supersymmetry
breaking $B$-terms,
$V = ...\,\, + B_{\alpha}(L^{\alpha}H_{2} + h.c.) + ...$, and 
$B_{\alpha} \propto F^{i}\partial_{i} \mu_{\alpha}$ (for example, see \cite{louisk}).
An interesting scenario arises 
in the case that $\mu_{\alpha} \propto \langle z_{1} \rangle^{n_{1}} 
\times \, .\, .\, .\, \times \langle z_{N} \rangle^{n_{N}}$
(we discuss below cases in which $\mu_{\alpha}$ is a mixture of 
non-vanishing $z$ and $F_{Z}$ components) with
$\langle z_{1} \rangle \ll \langle z_{j} \rangle$, where
$1 < j \leq N$, $n_{1} = 1$
and $\sum_{j}n_{j}R_{j} = 2 - R_{1}$ or  $1 - R_{1}$.
The parameter $B_{\alpha}$ is dominated by the 
$\partial_{i=1}\mu_{\alpha}$
contribution and could be of a different order
of magnitude than $\mu_{\alpha}$.

An interesting example is the case of $\mu = {\cal{O}}(1$ GeV)
and $B_{\alpha} = {\cal{O}}(M_{Z}^{2})$, i.e.,  
$\langle z_{1} \rangle \sim 10^{-2} \langle z_{j} \rangle$. 
Also, a small $\mu_{L}$ does not necessarily imply in this case
a small lepton number violation in the scalar potential.
[We comment on models with $\mu_{H} \sim \mu_{L} \sim  {\cal{O}}(1$ GeV)
(e.g., \cite{eg,halletal,fengetal}) below.]
Note that $B_{\alpha}$ are  the only lepton-number violating
soft supersymmetry breaking parameters in our models.

\subsection{A single $R =1$ superfield }

All three classes 
of $\mu$ parameters could be obtained from a single $R=1$ chiral superfield,
$Z$, with non-vanishing scalar and $F$ (auxiliary) components,
and whose fermion component, $\tilde{z}$,  condenses.
It may be  difficult to incorporate such a scenario in a realistic model.
Nevertheless, it is worth noting.
The different $\mu$ parameters are now given by
\begin{mathletters}
\label{e3}
\begin{equation}
\mu_{H} = \frac{\langle z \rangle^{2}}{M},
\label{e3a}
\end{equation}
\begin{equation}
\mu_{L} = \frac{F_{Z}^{*}}{M},
\label{e3b}
\end{equation}
\begin{equation}
\mu_{N} = \frac{\langle \tilde{z}\tilde{z} \rangle}{M^{2}},
\label{e3c}
\end{equation}
\end{mathletters}
and from (\ref{e3a}) one has $\langle z \rangle = {\cal{O}}(10^{11}$ GeV).
For $F_{Z}^{*} \sim (10^{11}$ GeV)$^{2}$  one has
$\mu_{L} \sim \mu_{H}$,
and  there is no clear hierarchy between the two parameters.
Lastly, if the condensate   $\langle \tilde{z}\tilde{z} \rangle \sim \Lambda^{3}$,
where $\Lambda \sim \sqrt{M_{SUSY}M} \sim 10^{11}$ GeV 
is the supersymmetry breaking scale,
then contrary to the generic situation
(see Example A), $\mu_{N} < \mu_{H}$.
As we discuss in section IV, the couplings of the right-handed
neutrino need to be suppressed in this case.
The relation (\ref{e3c}) would also allow for weak-scale vector-like exotics.
It is interesting to note that, regardless of its origin,
$\theta \sim 10^{11}$ GeV in all operators 
in this case. 

Regarding the Yukawa coupling $h^{\mbox{\tiny LNV}}$,
the holomorphicity of the superpotential forbids 
$h^{\mbox{\tiny LNV}} \sim z^{*}/M$. 
Thus,
one has  $h^{\mbox{\tiny LNV}} \lesssim |\mu_{L}/M| \rightarrow 0$.

If $Z$ was not an hidden-sector field, then the renormalizable 
superpotential term $ZLH_{2}$ would have been allowed
(leading to $\mu_{L} \sim 10^{11}$ GeV).
In such a case one needs to impose an additional symmetry, e.g.,
the usual $Z_{2}$ $R$-parity with $R_{P}(Z) = R_{P}(H_{1,\, 2}) = 
-R_{P}(L_{\tau,\, \mu,\, e}) = (+)$,
which allows for (\ref{e3a}) -- (\ref{e3c}) but forbids
the $R_{P}(ZLH_{2}) = (-)$ renormalizable term.

\subsection{$R =1/2$ and $R=-1/2$ superfields: Scalar vev's and F-terms}

In the case of a $R_{1} = 1/2$ superfield, 
$\mu_{L} = \langle z_{1} \rangle^{2}/M$
implies that $\langle z_{1} \rangle = {\cal{O}}(10^{11}$ GeV). Thus, 
$\mu_{H} = \langle z_{1} \rangle^{4}/M^{3} \rightarrow 0$ 
and like in Example A, the sneutrinos
could contribute significantly to electroweak symmetry breaking.

Having an additional field $Z_{2}$ with $R_{2} = -1/2$, and if the
$F_{Z_{1}} = {\cal{O}}(10^{11}$ GeV)$^{2}$ is non-vanishing,
one can obtain instead
\begin{mathletters}
\label{e4}
\begin{equation}
\mu_{H} = \frac{F_{Z_{1}}^{*}\langle z_{1} \rangle}{M^{2}},
\label{e4a}
\end{equation}
\begin{equation}
\mu_{L} = \frac{F_{Z_{1}}^{*}\langle z_{2} \rangle}{M^{2}},
\label{e4b}
\end{equation}
\begin{equation}
\mu_{N} = \frac{\langle z_{1} \rangle\langle z_{2} \rangle}{M}.
\label{e4c}
\end{equation}
\end{mathletters}
$\mu_{H,\,L}$  are  ${\cal{O}}(M_{Z})$ parameters in that case
if $\langle z_{i} \rangle \approx {\cal{O}}(M)$
(and there are two scale parameters $\theta_{1} \sim 10^{11}$ GeV
and $\theta_{2} \sim 10^{18}$ GeV,
as is often the case in supergravity models),
which implies that $\mu_{N}  \sim M$
[and again, there is no clear hierarchy between (\ref{e4a}) and (\ref{e4b})].

However, one has to forbid in that case
the $R = \pm 1$ combinations
$\langle z_{1,\,2} \rangle^{2}$, etc., which would lead to unacceptably large
$\mu_{L} = {\cal{O}}(M)$ and $h^{\mbox{\tiny LNV}} = {\cal{O}}(1)$.
For example, 
a continuous hidden sector $U(1)$ symmetry
with identical charge assignments to both fields would allow
(\ref{e4a}) and (\ref{e4b}) but forbid the dangerous couplings
(as well as $\mu_{N}$).

\subsection{Lessons and comments}

Our main lesson is that the typical situation
$\mu_{N} \sim M$ and 
$h^{\mbox{\tiny LNV}} \rightarrow 0$ is easily found
in simple examples of $U(1)_{R}$ selection rules
(but exceptions exist).
If $\mu_{\alpha} \propto \prod_{i} \langle z_{i}^{n_{i}} \rangle$
then typically $\mu_{H} \lesssim \mu_{L}$. However,
if there are also non-singlet $F$-terms which do not vanish,
then no clear hierarchy exists.
In some cases additional symmetries may be required
in order to suppress
$h^{\mbox{\tiny LNV}}$,
but typically $U(1)_{R}$ is sufficient for that purpose.
(The additional symmetries, if required,  could correspond to 
typical symmetries that are often found in models.)

We presented a few scenarios in which some, or all,
of the $\mu_{H,\, L,\, N}$ are generated.
Many more scenarios exist.
The relevant scenario would be determined
by the realization of the $U(1)_{R}$ symmetry,
the hidden singlet $R$-charges, and by the relations
between the fields that spontaneously break $U(1)_{R}$
and supersymmetry breaking.

If the SM fields do not have the standard $R$-charge assignments 
\cite{gaugeR} that we assume, then some of the examples given above
may need to be revised, depending on the charge assignments chosen.
If it is a family-dependent assignment then
one could distinguish (unlike in our case)
between  $\mu_{L_{\tau}}$, $\mu_{L_{\mu}}$,  and $\mu_{L_{e}}$.
(We discuss family dependences in Section VII.)

Lastly, if there is a hidden $R = 0$ chiral superfield $Z$
with $F_{Z} =  {\cal{O}}(10^{11}$ GeV) then $\mu_{H}$
(but not\footnote{Note that if $\mu$ is generated in this manner
then it is proportional to $W$ and has $R$-charge $R(\mu)= 2$.} 
$\mu_{L}$) could be partially generated
by a Kahler potential source \cite{gm}, smearing any
correlation between $\mu_{H}$ and $\mu_{L}$.

\section{Conditions for  light (neutrino) eigenstates}
\label{sec:s3}

The ratio $\mu_{L}/\mu_{H}$, which is a (high-energy) order parameter 
of the models,  
need not be suppressed
in order to suppress the tree-level neutrino mass.
This is a trivial result of the observation that there are
no tree-level masses for the neutrinos if 
$\mu_{\alpha}$ and $\langle L_{\alpha}^{0} \rangle$
are aligned in field space \cite{hs}. (This observation is also
the basis for the recent works
of Refs.\ \cite{bn,ralf}.)
We elaborate on that observation in this section,
and we show in the next section that
nearly parallel $\mu_{\alpha}$ and $\langle L^{0}_{\alpha} \rangle$
four-vectors arise dynamically in simple models
for the soft supersymmetry breaking parameters (dynamical alignment).
Discussion of laboratory and cosmological implications 
is postponed to sections V and VI, respectively.

An important distinction 
(which allows for the neutrino mass generation)
between the models that are discussed here and the MSSM 
is that all seven neutral fermions mix in our case.
That is, the supersymmetric partners of the $B$ and $W_{3}$ gauge bosons 
and neutral Higgs --
the bino, wino (the gauginos) and the two Higgsinos, respectively  (the neutralinos),
mix with each other and with
the three neutrinos once elelectroweak symmetry is broken.
Hence, one has a seven-dimensional Majorana mass matrix\footnote{
All relevant mass and mixing matrices are given, 
for example, in Ref.\  \cite{ralf}.}
for the neutral fermions, which involves (at tree-level)
\begin{itemize}
\item
the bino and wino soft supersymmetry breaking masses
$M_{1}$ and $M_{2}$, respectively, 
\item
the Higgsino and neutrino masses
$\mu_{\alpha}$, 
\item
and the gauge-matter mixing parameters
$M_{Z}\nu_{2}/\nu$ and  $M_{Z}\nu_{\alpha}/\nu$. 
\end{itemize}
(We define
$\nu_{2} = \langle H_{2}^{0} \rangle$,
$\nu_{\alpha} = \langle L_{\alpha}^{0} \rangle$,
$\nu_{1} = |\nu_{\alpha}|$, $\tan\beta = \nu_{2}/\nu_{1}$, 
and $\nu = \sqrt{\nu_{2}^{2} + \nu_{1}^{2}}$.) 
The determinant of the neutral fermion mass matrix vanishes:
The $SU(4)$ symmetry is broken down to $SU(2)$ 
by two fundamental vectors, $\mu_{\alpha}$ and $\nu_{\alpha}$, 
each of which can render only one physical state massive,
and there are two massless states. 
One could define, for example, the $e$ and $\mu$ neutrinos
as  those  states,
$m_{\nu_{e}} = m_{\nu_{\mu}} = 0$.

Let us then eliminate the (tree-level) massless states and 
discuss the residual five-dimensional non-degenerate mass matrix.
It is convenient to define for that purpose, following Banks et al.\ \cite{bn}, 
$\cos\zeta \equiv \langle \mu_{\alpha}\nu^{\alpha} \rangle /
 |\mu_{\alpha}||\nu_{\alpha}|
\equiv \langle \mu_{\alpha}\nu^{\alpha} \rangle / \mu\nu_{1} $.
The determinant (of the residual mass matrix) 
is proportional to $\sin^{2}\zeta$, which is the second
(or the low-energy)
order parameter of the model. 
That is, if $\mu_{\alpha}$
and $\nu_{\alpha}$ are aligned [and $SU(4)$ is broken down to only $SU(3)$] 
then only one state is rendered massive and there is an additional
massless state $m_{\nu_{\tau}} = 0$.
Otherwise, the lightest neutral fermion, which we
define to be the $\tau$ neutrino $\nu_{\tau}$, has 
a mass $m_{\nu_{\tau}}  \sim M_{Z}\sin^{2}\zeta$
(assuming, for simplicity, $M_{SUSY} \sim M_{Z}$). It is
sufficiently suppressed and phenomenologically acceptable
if $\sin\zeta \lesssim \sqrt{m_{\nu_{\tau}}^{\mbox{\tiny exp}}/M_{Z}}$
(where $m_{\nu_{\tau}}^{\mbox{\tiny exp}} \sim 23$ MeV [$100$ eV]
is the laboratory \cite{aleph} [cosmological\footnote
{The $\sim 100$ eV bound applies to stable neutrinos
and assumes the critical energy density $\Omega h^{2} = 1$.
If the neutrinos constitute only a part of that energy density, 
then the upper bound scales down accordingly, i.e.,
$m_{\nu} \lesssim 100 \Omega_{\nu} h^{2}$ eV.} 
(energy density) \cite{100ev}] upper bound on the neutrino mass).

The alignment condition and all other sufficient conditions
for the suppression of the neutrino mass
are more easily seen in the rotated basis in which
$\mu_{\alpha} = \mu(1,\, 0)$ (assuming one lepton generation).
One has\footnote{The determinant of the MSSM neutralino mass matrix
reads $\det M_{\mbox{\tiny neutralino}} 
= \mu \nu_{1}\nu_{2}(g_{1}^{2}M_{2} + g_{2}^{2}M_{1})
- \mu^{2} M_{1}M_{2}$. The tree-level neutrino mass
is given by the ratio 
$\det M_{\mbox{\tiny neutral}}/\det M_{\mbox{\tiny neutralino}}$.}
(for the residual mass matrix)
in that basis 
\begin{equation}
\det M_{\mbox{\tiny neutral}} = \mu^{2}\tilde{\nu}_{L_{\tau}}^{2}(
g_{2}^{2}M_{2} + g_{1}^{2}M_{1}),
\label{det}
\end{equation}
where $g_{1,\, 2}$ are the hypercharge and $SU(2)$ couplings,
respectively, and the rotated 
$\tilde{\nu}_{L_{\tau}} \approx 
\nu_{L_{\tau}}[1 - (\mu_{L_{\tau}}\nu_{H_{1}}/
\mu_{H_{1}}\nu_{L_{\tau}})]
\approx
\mu_{L_{\tau}}[(\nu_{L_{\tau}}/\mu_{L_{\tau}}) - (\nu_{H_{1}}/\mu_{H_{1}})]$ 
(assuming $\mu_{H_{1}} > \mu_{L_{\tau}}$).
Thus, there are light states if \cite{eg} 
\begin{itemize}
\item $(a)$  $\tilde{\nu}_{L_{\tau}} \rightarrow 0$, i.e., 
$\mu_{\alpha}$ and $ \nu_{\alpha}$ are parallel
(and $\sin\zeta \rightarrow 0$) or $\tan\beta \gg 1$
(and $\nu_{1} \rightarrow 0$), 
\item $(b)$ $\mu \rightarrow 0$, 
\item and $(c)$
the gauginos are heavy and their mixing 
with the matter fermions is negligible ,
i.e., $M_{1},\, M_{2} \gg \mu,\, M_{Z}$, in which 
case $ \{ \det M_{\mbox{\tiny neutral}}\}_{\mbox{\tiny matter}} \sim
\det M_{\mbox{\tiny neutral}}/M_{1}M_{2}$ (see above footnote).
\end{itemize}
It is, however, difficult to realize the latter case because
$\mu$ and the gaugino masses both contribute
to the quadratic terms in the scalar potential, and thus, are related by the 
minimization conditions described below
and are typically of the same order of magnitude.
 
The case $(b)$ is quite interesting and was discussed in Refs.\ \cite{eg,halletal}.
(See also Ref.\ \cite{fengetal}
for a recent discussion and references 
of the equivalent situation in the MSSM.
However, the constraints on the parameter space in our
case could be different than in the MSSM.)
At least two of the neutral fermions, a Higgsino and a neutrino,  are now light.
The light Higgsino contribution to the $Z$ width constrains 
the model to the region of $\tan\beta \sim 1$.
From RSB one has that the scalar mixing parameter 
$B_{\alpha}$ (see Section IV)
is not proportional to $\mu_{\alpha}$ and cannot vanish
in that case.
Thus, the  Higgs boson mass would partially 
come from mixing ($\propto B_{\alpha}$)
with the scalar neutrinos.
To consistently realize such a scenario would require a departure
from the simple models for the soft terms that we consider below.
We will comment on that scenario again when discussing
the scalar potential, but we do not consider it in detail.

Perturbativity of Yukawa couplings and dynamical alignment 
(which are considered in Section IV) constrain
$\tan\beta$  from above (see below).
Thus, the remaining case $(a) $ is typically realized for
nearly aligned vectors, i.e.,  $\tilde{\nu}_{L_{\tau}} \rightarrow 0$.
(This is crucial for the discussion of the phenomenological implications
of the model in sections V and VI.) Nevertheless,
there are additional (secondary) suppressions of $m_{\nu}$
if $\tan\beta \gtrsim 2-3$ or
if the gaugino mass parameters are large.
For example, in Fig.\ 1 we show $m_{\nu_\tau}$ as a function of $\sin\zeta$
for different values of the gaugino masses and of $\tan\beta$ 
(but no RSB is yet required). Note, in particular, the high precision
required in the alignment for moderate values of $\tan\beta$.
The alignment condition, however, can be slightly relaxed
for large values of $\tan\beta$.


\begin{figure}[t]
\label{fig:fig1}
\postscript{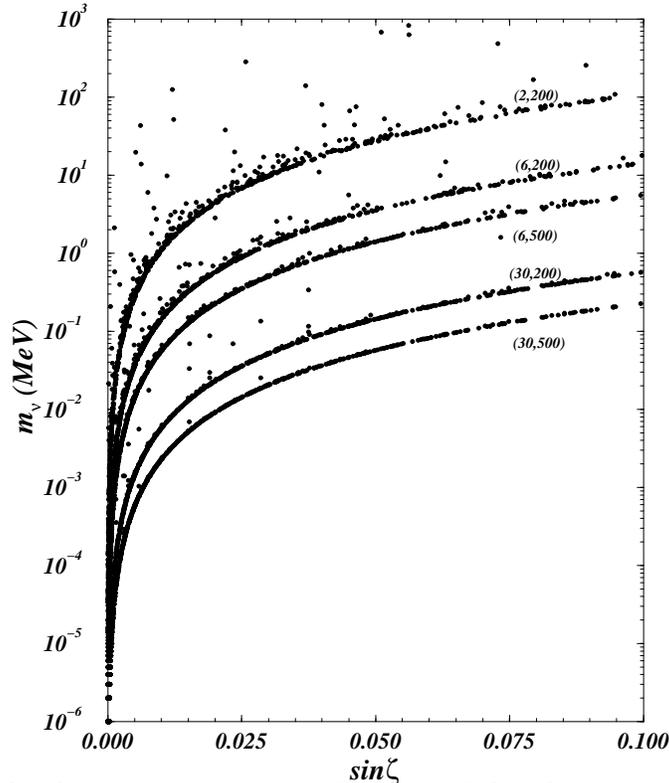}{0.5}
\caption{
The tree-level neutrino mass as a function of the alignment (low-energy)
order parameter $\sin\zeta$ for $(\tan\beta,M_{2}/$GeV) = (2,200), (6,200), (6,500),
(30,200), (30,500). For simplicity we assume the gaugino-unification
relation $M_{1}/M_{2} = 5/3 \times g_{1}^{2}/g_{2}^{2}$. The $SU(4)$ directions 
and $\mu$ are taken to be free parameters. No radiative symmetry breaking is yet
required.
}
\end{figure}

We consider in this work only the tree-level mass matrix.
The one-loop matrix has
new non-diagonal entries [that explicitly break the residual
$SU(2)$ symmetry] and 
all seven states are massive with generically 
one very light neutrino\footnote{The Yukawa origin of this
mass is the same as in the models of Ref.\ \cite{yukawaloop1,yukawaloop2}.
Since renormalization effects are neglected there, they
have $\mu_{L} \equiv 0$ and no tree-level or $D$-vertex one-loop
masses as in the general case.} 
and one neutrino with a mass 
$\sim {\cal{O}}(10^{-3} - 1 \times m_{\nu_{\tau}})$ 
(i.e., it could be of the order of the tree-level mass)
\cite{hs,dawson,ralf}.
Thus, even though two neutrinos are massless here,
all neutrinos are massive in a more complete calculation
(which is not necessary for our purposes).
We would also like to point out that in four-family models, which 
are constrained by the $Z$-width to have
$m_{\nu_{4}} \gtrsim 45$ GeV, $\sin\zeta$ is not constrained by the tree-level 
neutrino mass (which could partially account for $m_{\nu_{4}}$)
but by the size of the 
corresponding loop masses for the ordinary neutrinos
(as well as by other phenomenological considerations).

\section{The scalar potential and dynamical alignment}
\label{sec:s4}

In a general lepton number violating extension of the MSSM
the scalar neutrinos (sneutrinos)
will typically acquire  vev's so that lepton number is broken 
both explicitly and spontaneously.
Containing  all lepton number violation
in $W_{M}$ (as in our case) is, however,  sufficient to spontaneously
break lepton number. (In this section lepton number refers
to the high-energy lepton number.)
Below,  we will organize the Higgs-lepton scalar potential
and its minimization equations in a convenient way, and show that
indeed the sneutrinos aqcuire vev's. We would then list the conditions
for achieving the required dynamical alignment between
$\mu_{\alpha}$ and $\nu_{\alpha}$, and show that they are satisfied
in a large class of models, but not in generic
grand-unified theories and in models with an intermediate-scale
right-handed neutrino.
We will also comment on perturbativity constraints
on the Yukawa couplings (which are different than in the MSSM).
For simplicity, we will discuss  models
with only the third generation of quarks and leptons.
The generalization, however,  is straightforward.
(Renormalization-group scaling of general broken $R_{P}$ extensions of the MSSM
were discussed recently in Ref.\ \cite{rges}.)

It is convenient to define
\begin{mathletters}
\label{defs}
\begin{equation}
L_{\alpha} = L(\cos\alpha,\, sin\alpha),
\label{defsl}
\end{equation}
\begin{equation}
\mu_{\alpha} = \mu(\cos\gamma,\, sin\gamma),
\label{defsmu}
\end{equation}
\begin{equation}
B_{\alpha} = B(\cos\delta,\, sin\delta),
\label{defsb}
\end{equation}
\end{mathletters}
and $\tan\beta = \nu_{2}/|\langle L^{0} \rangle |  =
\nu_{2}/\nu_{1}$ as before. (Note that $B_{\alpha}$ has a squared mass
dimension.)
The selection rules of Section II 
typically imply lepton number
conservation by dimensionless couplings 
of the low-energy Kahler potential.
Thus, the scalar masses $m^{2}_{\alpha\beta}L^{\alpha}L^{\beta *} 
= \mu_{\alpha}\mu_{\beta}^{*}L^{\alpha}L^{\beta *}$ 
and $B_{\alpha}L^{\alpha}H_{2}$
are the only
source of explicit lepton number violation in the scalar potential.
(The Yukawa $F$ terms must involve, in our case,
charged degrees of freedom, and thus, vanish at the minimum.)
Hence, the Higgs-lepton scalar potential can 
be written as a straightforward generalization of the MSSM
Higgs scalar potential, i.e.,
\begin{equation}
V(L_{\alpha},\, H_{2}) 
= m_{1}^{2}L^{2} 
+ m_{2}^{2}H_{2}^{2} + m_{3}^{2}(LH_{2} + h.c.) + 
\frac{1}{8}(g_{1}^{2} + g_{2}^{2})(H_{2}^{2} - L^{2})^{2} + \Delta V,
\label{potential}
\end{equation}
where $\Delta V$ is the one-loop correction (that is included in
our numerical procedures\footnote{We include
only corrections proportional to the $t$ and $b$ Yukawa couplings,
which in our case, are given by the 
corresponding corrections in the MSSM.}) and 
\begin{mathletters}
\label{defs2}
\begin{equation}
m_{1}^{2} = m_{H_{1}}^{2}\cos^{2}\alpha + m_{L_{\tau}}^{2}\sin^{2}\alpha
+ \mu^{2}\cos^{2}(\alpha - \gamma),
\label{m1}
\end{equation}  
\begin{equation}
m_{2}^{2} = m_{H_{2}}^{2} + \mu^{2},
\label{m2}
\end{equation}
\begin{equation}  
m_{3}^{2} = B\cos(\alpha - \delta),
\label{m3}
\end{equation}
\end{mathletters}
where on the right-hand side $m_{i}^{2}$ 
is the soft supersymmetry breaking squared mass
of scalar $i$.

Note that the (tree-level)
scalar potential (\ref{potential}) has, in the absence of explicit
lepton number breaking ($\gamma = \delta = 0$),
the unbounded direction
$m_{L_{\tau}}^{2} + (1/8)(g_{1}^{2} + g_{2}^{2})(\langle H_{2}^{0}\rangle^{2} 
- \langle H_{1}^{0}\rangle^{2})^{2} < 0$
\cite{aulakh},
as well as a flat direction 
$\langle L^{0} \rangle^{2} - \langle H_{2}^{0} \rangle^{2} = 
-4m_{L_{\tau}}^{2}/(g_{1}^{2} + g_{2}^{2})$ \cite{flat,bs,Majoron}.
The MSSM $D$-flat direction
$\langle L^{0} \rangle^{2} - \langle H_{2}^{0}\rangle^{2}  = 0$ 
(i.e., $\tan\beta = 1$) is also relevant here and could also lead
to a flat direction (one could eliminate $m_{L_{\tau}}^{2}$ in that case
by redefinitions of $m_{H_{1}}^{2}$ and $m_{H_{2}}^{2}$).
Lepton number could
be spontaneously broken in the case of a  flat direction, even if it is 
not broken explicitly. This possibility was discussed
in the context of supersymmetric Majoron models \cite{aulakh,eg,rv,flat,bs,Majoron}.
We will exclude these directions, which are difficult to realize
once RSB is included, from our analysis.

It is convenient to minimize 
$V(L_{\alpha},\, H_{2})$
with respect to $\langle L^{0} \rangle = \nu_{1}$, 
$\langle H_{2}^{0} \rangle = \nu_{2}$, and the angle $\alpha$.
Two of the minimization equations reduce to those of the MSSM, i.e.,
\begin{equation}
\frac{m_{1}^{2} - m_{2}^{2}\tan^{2}\beta}{\tan^{2}\beta -1}
=
\frac{1}{2}M_{Z}^{2},
\label{min1}
\end{equation}
\begin{equation}
m_{3}^{2} = -\frac{1}{2} \sin 2 \beta \left[ m_{1}^{2} + m_{2}^{2} \right].
\label{min2}
\end{equation}
In particular, At the minimum $V = V^{\mbox{\tiny MSSM}}  = 
-\frac{1}{8}M_{Z}^{2}\nu^{2}\cos^{2} 2 \beta$, independent of
$\alpha,\, \gamma,\, \delta$.
The third equation reads
\begin{equation}
\left[(m_{H_{1}}^{2} - m_{L_{\tau}}^{2})\sin 2\alpha + \mu^{2}
\sin 2 (\alpha - \gamma) \right] \cos\beta
+2B\sin (\alpha - \delta)\sin \beta  = 0.
\label{min3}
\end{equation}

For the boundary conditions 
\begin{mathletters}
\label{bc}
\begin{equation}
m_{H_{1}}^{2} = m_{L_{\tau}}^{2},
\label{bc1}
\end{equation}
\begin{equation}
\gamma = \delta,
\label{bc2}
\end{equation}
\end{mathletters}
there is only one $SU(4)$
direction in field space which is determined by  the angle $\gamma$.
Hence, $\nu_{\alpha}$ must align along that direction and 
the solution to (\ref{min3}) (for $\mu \neq 0$, $B \neq 0$,
and away from the flat direction)  is given by $\alpha = \gamma$.
The alignment is achieved dynamically in that case.
Note that the $\mu_{\alpha}$ is a parameter vector in the low-energy theory
and the $SU(4)$ symmetry is broken explicitly and spontaneously down  
to (the low-energy lepton symmetry) $SU(3)$, and (the rotated low-energy) $\tilde{H}_{1}$
contains the seven pseudo Goldstone bosons with mass $\propto \mu$.
Unlike models with no explicit breaking (but only
spontaneous breaking along the flat direction) there are no
light Majorons.

The boundary conditions (\ref{bc}) are often found in models 
for the soft supersymmetry breaking terms.
The second condition is trivially realized if $B_{\alpha} = b\mu_{\alpha}$
(i.e., $B$-proportionality), as is also often the case.
The crucial point (which was also noted independently
by Hempfling \cite{ralf}) is that if 
$m_{H_{1}}^{2} = m_{L_{\tau}}^{2} = m_{0}^{2}$
and $B_{\alpha} = b\mu_{\alpha}$
at some Planckian scale, then the deviations from these relations
at the weak scale due to renormalization group scaling
are proportional to the square of the $b$ Yukawa coupling $h_{b}$, i.e.,
\begin{equation}
\frac{\partial m_{H_{1}}^{2}}{\partial \ln Q} = 
\frac{\partial m_{L_{\tau}}^{2}}{\partial \ln Q}  
+ \frac{h_{b}^{2}}{8\pi^{2}}[m_{H_{1}}^{2} + 
m_{Q_{3}}^{2} + m_{D_{3}}^{2} +A_{b}^{2}],
\label{rge}
\end{equation}
where $A_{b}$ is the trilinear soft mass parameter $A_{b}H_{1}Q_{3}D_{3}$. 
Thus, deviations from universality are strongly suppressed (unless $\tan\beta \gg 1$, i.e.,
$h_{b} \sim 1$, or if the soft parameters $m_{H_{1}}^{2} + 
m_{Q_{3}}^{2} + m_{D_{3}}^{2} +A_{b}^{2} \gg  m_{L_{\tau}}^{2} $),
but are still sufficient to generate the small neutrino mass.

This is the case in the (extended) MSSM with the so-called
universal boundary conditions
at the grand scale (i.e., $m^{2}_{\mbox{\tiny scalar}} \equiv m_{0}^{2}$,
$A_{i} \equiv A_{0}$, and
$M_{\mbox{\tiny gaugino}} \equiv M_{1/2}$).
We solved the system (\ref{min1}) -- (\ref{min3}) iteratively\footnote{
Given $\beta$ and $\gamma$, we use Eq.\ (\ref{min1}) to solve for $\mu$,
Eq.\ (\ref{min2}) to solve for the proportionality parameter $b$, 
and Eq.\ (\ref{min3}) to solve (numerically) for $\alpha$.}
in our model for various universal
boundary conditions and confirmed the above assertions and claims.
We also included in the numerical
procedures the one-loop radiative corrections (see above)
and the corrected $b$ and $\tau$ Yukawa couplings (see below).

Unlike in the MSSM, $m_{b} \neq h_{b}\langle \tilde{H}_{1}^{0} \rangle
= h_{b}\nu\cos\beta$,
but rather $m_{b} = h_{b}\langle{H}_{1}^{0} \rangle =
h_{b}\langle \tilde{H}_{1}^{0} \rangle \cos\alpha$, 
leading to the weak-scale
perturbativity constraint (requiring $h_{b} < 1$, i.e., that it is below its
quasi fixed-point)
\begin{equation}
\cos \alpha \cos \beta  > \frac{m_{b}}{174\, {\mbox{GeV}}} \sim \frac{1}{58}.
\label{hb}
\end{equation}
It can also be written as $\tan\beta < 58\cos\alpha$. 
Eqs. (\ref{rge}) and (\ref{hb}) imply that the models
are realized more naturally for small and moderate
values of $\tan\beta$ (while typical grand-unified models
with right-handed neutrinos are realized for large $\tan\beta$ \cite{smir}).
Phenomenological implications,
which are  discussed in the following sections, constrain
$\tan\gamma \sim \tan\alpha$ from above, i.e., $\tan\alpha \lesssim 1$, 
which is stronger
than any constraint that one could derive from (\ref{hb}).
Also, unlike in the MSSM, the leptonic yukawa couplings
$h_{\tau,\, \mu,\, e} \neq m_{\tau,\, \mu,\, e}/ \nu\cos\beta$
but are found by requiring the three light eigenstates
of the charged (color singlet) fermion mass matrix
to have the correct masses, i.e., mixing effects
can slightly modify their values.
[Since the leptonic Yukawa couplings have nearly
flat renormalization curves (for not too large $\tan\beta$),
this effect could be important when considering relations
between Yukawa couplings at Planckian scales.]
We account for these effects in our numerical procedures. 


\begin{figure}[t]
\label{fig:fig2}
\postscript{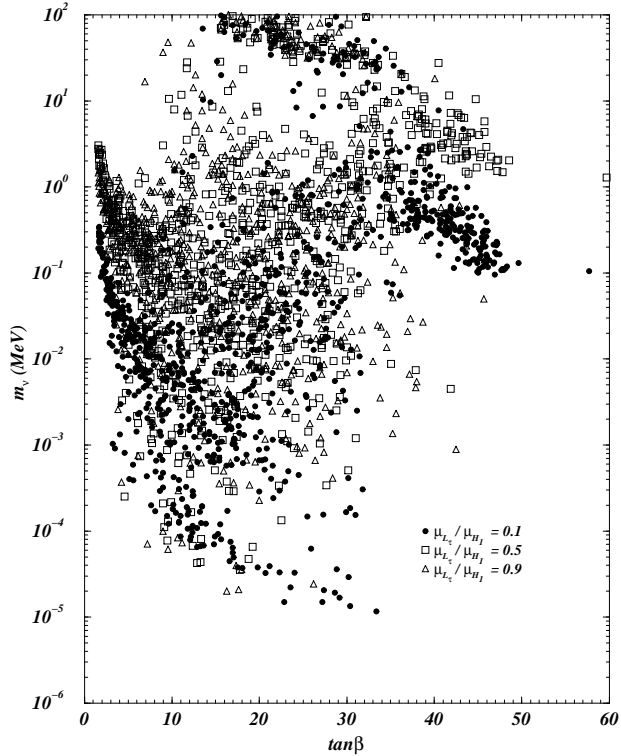}{0.5}
\caption{
The allowed parameter space 
(imposing cuts from direct searches, RSB,  and stability considerations)
for $0 \leq m_{0} \leq 1000$ GeV, $|A_{0}/m_{0}| \leq 3$,
and $70 \leq M_{1/2} \leq 500$ GeV, is scanned, and the prediction for the neutrino mass
is calculated and shown as a function of $\tan\beta$.
We assume $\tan\gamma = \mu_{L}/\mu_{H} = 0.1$ (bullets), 0.5 (squares), and 0.9 (triangles).
(For definiteness, the $t$-quark pole mass is $m_{t}^{pole} = 175$ GeV.)
}
\end{figure}


In Fig.\ 2 we show the neutrino mass as a function of $\tan\beta$
for $\tan\gamma = 0.1,\, 0.5,\, 0.9$.
The corresponding alignment, i.e., comparison of (the output) $\tan\alpha$ 
and (the input) $\tan\gamma$ is shown in Fig.\ 3
for $\tan\gamma = 0.1$ (bullets) and 0.9 (squares).
In Fig. 4 we further explore the parameter space for $\tan\gamma = 0.1$.
By observation, we can draw the following conclusions, some of which
are somewhat surprising:
\begin{itemize}
\item
The typical suppression factor is of ${\cal{O}}(10^{-5})$, 
i.e., $m_{\nu} \sim {\cal{O}}(1$ MeV). However, for
$\tan\beta \gtrsim 5$ one finds suppression as strong as 
${\cal{O}}(10^{-9})$ [or  even ${\cal{O}}(10^{-10})$], i.e.,
$m_{\nu} \sim {\cal{O}}(10 -100$ eV).
The functional dependence on the soft parameters is complicated.
In general, we find that $m_{0} \gg M_{Z}$ and 
$M_{1/2} < m_{0}$ are slightly prefered. 
The former leads to heavy scalars which are less
sensitive to the renormalization-group corrections, while the latter
suppresses the ${\cal{O}}(h_{b}^{2})$ contributions in (\ref{rge})
[which is a more important  consideration than the large 
$M_{2} \sim 0.8 M_{1/2}$ effect
in (\ref{det})]. 
\item
The smaller values of $\tan\gamma$ do not imply necessarily smaller
neutrino masses and vice versa.
\item
The alignment condition is indeed satisfied poorly for 
$\tan\beta \gg 1$, in contradiction
to its excellent satisfaction for small and intermediate values of $\tan\beta$.
This leads to a relatively wide (narrow) range for 
$m_{\nu}$ in the former (latter) case.
\item
The poor alignment  for $\tan\beta \gg 1$ 
and the suppression of $\nu_{1}$ are competing effects which
allow to realize small neutrino masses for large $\tan\beta$ in some cases.
\end{itemize}


\begin{figure}[t]
\label{fig:fig3}
\postscript{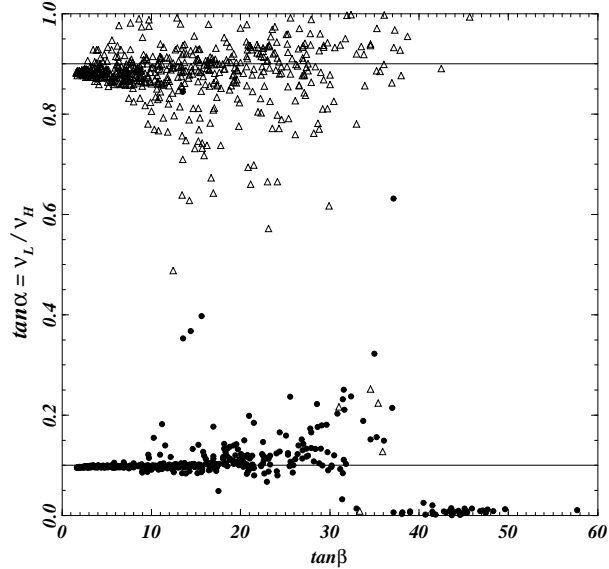}{0.5}
\caption{
Same as in Fig. 2 except for $\tan\alpha = \nu_{L}/\nu_{H}$
is shown for $\tan\gamma = 0.1$ (bullets) and 0.9 (triangles).
For comparison, $\tan\gamma = 0.1,\, 0.9$ is also indicated. 
}
\end{figure}

\begin{figure}[t]
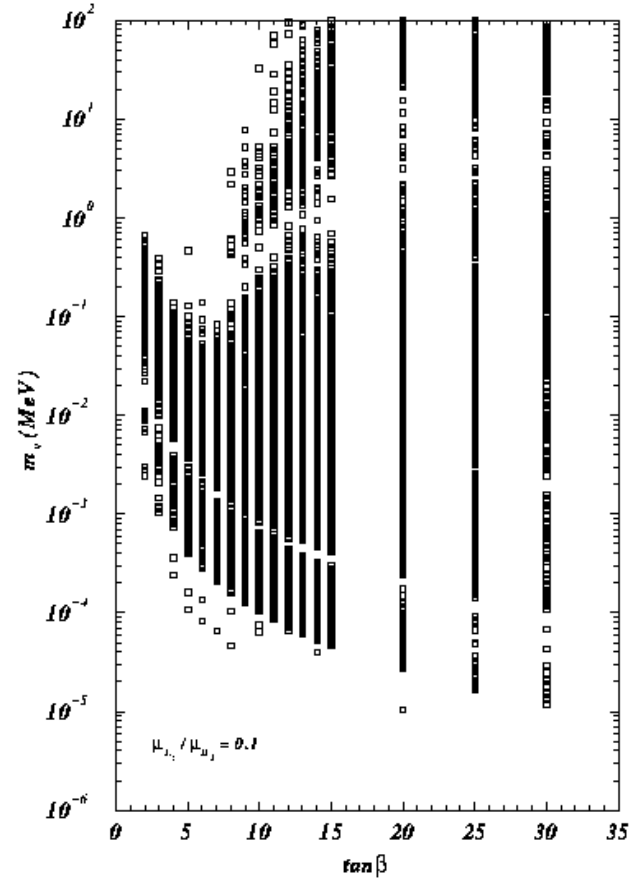

\label{fig:fig4}
\caption{
Same as in Fig. 2 except for $\tan\gamma = 0.1$
and the neutrino mass
is calculated and shown for discrere choices  of $2 \leq \tan\beta \leq 30$.
}
\end{figure}


Let us stress that our results hold in a larger class of models,
i.e., in models with universality of the $L_{\alpha}$ soft masses and
the proportionality $B_{\alpha} \propto \mu_{\alpha}$.
The lepton-Higgs universality is guaranteed if the Kahler potential is 
(approximately) invariant under the $SU(4)$, e.g., in string models
it would require that all $L_{\alpha}$ components have a unique set
of modular weights. No universal boundary conditions for 
any other fields are required. In grand-unified models, however,
the soft parameters are scaled at 
Planckian scales according to renormalization
group equations which are invariant under the grand-unified symmetry
and which are determined by its superpotential couplings.
In generic models the leptonic and Higgs fields are embedded
in different representations
(often with different dimensions)
and have different Yukawa interactions with  sometimes large
Yukawa couplings.
For example, $H_{1}$ would typically couple to the large Higgs
representation that break the unified symmetry and/or
render its SU(5) partner, the color triplet, heavy.   
Thus, the Higgs and lepton soft masses are subject to different scaling laws
at large scales. They could also receive different contributions from
$U(1)$ $D$-terms\footnote{If the $U(1)_{R}$ 
is gauged the respective (residual)
$D$-terms could lead to Higgs-lepton non universality.
Similar problems would also arise in the case of low-energy
stringy $U(1)$'s, unless the $D$-terms vanish or are negligible.
Thus, global symmetries are better suited for our purposes.} 
if the grand
unified symmetry is broken to some group $G \times U(1)$
[which is not the hypercharge $U(1)$] 
between the Planck and unification scales.
The $SU(4)$ symmetry would be broken in $V(L_{\alpha},\, H_{2})$
in more that one direction and $\nu_{\alpha}$
and $\mu_{\alpha}$ would not be aligned.
For example, we find that in the $SU(5)$ models of Ref.\ \cite{alex}
the scaling violation of the $SU(4)$
due to the $\lambda H_{1}\Sigma_{24}H_{2}$ term destroys, in most cases,
the dynamical alignment.

Similarly, a right-handed neutrino, $N$,  will introduce the Yukawa terms
$h_{N}H_{2}LN$, and unless $h_{N} \ll 1$ 
for all three generations 
(but in unified models typically  $h_{N} \approx h_{U}$)
or $\mu_{N} \approx M$,
it would again  introduce new $SU(4)$ breaking.
Thus, it is difficult to reconcile supersymmetric neutrino masses 
(with dynamical alignment)
and right-handed neutrino see-saw models.
(Similar arguments hold for an arbitrary high-energy 
$W_{Y}^{\mbox{\tiny LNV}}$.)

In general models for the soft terms one could always 
tune the parameters to satisfy 
(\ref{min3}). For example, if $\mu = {\cal{O}}(1$ GeV) 
then $m_{3}^{2} = {\cal{O}}(M_{Z}^{2})$ may not be proportional to $\mu$
(see Example B of Section II). It could lead to large mixing in the scalar sector,
but would require some tuning of parameters
(or to invoke extended symmetries)
in order to still guarantee $m_{\nu} \ll \mu$.
We do not find this possibility attractive and do not pursue it here.

\section{Collider and laboratory implications}
\label{sec:s5}

Having established the smallness of the neutrino mass
and its consistency with the laboratory bound 
$m_{\nu_{\tau}} \lesssim 23$ MeV \cite{aleph} 
(and in many cases with the cosmological bound 
$m_{\nu_{\tau}} \lesssim 100$ eV \cite{100ev}) in the previous sections,   
we will not discuss other phenomenological aspects of the models in detail
but only survey possible constraints. 
We will show that in our framework the constrained
couplings and amplitudes
are proportional to some power of
$m_{\nu_{\tau}}/M_{Z}$ or to small (MSSM) Yukawa couplings, 
and thus, are evaded in large sections
of the parameter space.
In particular, 
due to the large number of parameters,
it is difficult to efficiently constrain the models
(until  supersymmetric particles and
their decays are observed and characterized):
See, for example, Ref.\ \cite{constraints}
for detailed discussions of constraints\footnote{
Note that any constraints imposed there on
$\nu_{L_{\tau}}$ etc.,  apply, in our case, only to the rotated vev's
after the SM lepton number is defined (see below).}
from rare decays, weak current universality, etc.,
on these and on more general models of lepton
number violation.

The discussion is simplified once again if one performs a 
low energy $SU(4)$ rotation. 
For our  purposes here it is convenient to define  
the low-energy Higgs field $\tilde{H}_{1}$ such that
$\langle \tilde{H}_{1} \rangle = \nu_{1}$, and the three leptonic fields
$\tilde{L}_{\tau,\, \mu,\, e}$ correspond to the three
perpendicular directions in field space.
In the degenerate limit $\sin\zeta \rightarrow 0$ one then has
$\mu_{\alpha}L^{\alpha}H_{2} \rightarrow  \mu\tilde{H}_{1}H_{2}$.
All fermion and scalar mass matrices are block diagonal
in the gauge-(low-energy) Higgs fields and in the (low-energy) leptonic
fields.   
The rotated neutral and charged states constitute the ``physical'' lepton
and Higgs doublets consistently, and one can define the (low-energy)
SM  lepton number
in this basis. (If $\mu_{L} \gtrsim \mu_{H}$ 
then the discrepancy between low and
high energy definition of leptons is maximized, leading to possible
constraints on their ratio.)
The rotation generates
$W_{Y}^{\mbox{\tiny LNV}}$ [see (\ref{wlnv})] and thus,
$W_{Y} \neq W_{Y}^{\mbox {\tiny MSSM}}$ and $W$ has an accidental
$Z_{3}$ (baryon parity \cite{z3}) symmetry 
(but only at the renormalizable level).

Since $\sin\zeta$ is small but not exactly zero
(the alignment, and in particular, if achieved dynamically,
is not scale invariant and is not expected to be exact)
there are perturbations to the naive limit, and some mixing between the 
(low-energy) leptonic and gauge-Higgs sectors arises, e.g., the $\tau$
could contain a small admixture of the wino and the Z could
have off-diagonal couplings to a neutrino and a gaugino.
However, the effects are of the order of magnitude
of the neutrino mass suppression factor\footnote{
For large $\tan\beta$, however, $\nu_{1} \rightarrow 0$
allows for larger values of $\sin\zeta$. In that case, experimental
constraints could be effective in constraining the models.} 
$\sin\zeta \sim \sqrt{m_{\nu_{\tau}}/M_{Z}}$ to some power.
Note that since 
one of the loop-induced neutrino masses
could be of the order of the tree-level mass, 
experimental constraints should be applied to
the mixing parameters extracted from
the one-loop mass matrices.

Other constraints result  from 
the new Yukawa interaction contributions   to
flavor changing neutral (FCNC)
and charged currents.  The superpotentials
$W_{Y}^{\mbox{\tiny LNV}}$
and $W_{Y}^{\mbox{\tiny MSSM}}$ are related by a rotation and 
$W_{Y}^{\mbox{\tiny LNV}} \propto W_{Y}^{\mbox{\tiny MSSM}}$.
As a result,  $h_{D}$ and $h_{D}^{\mbox{\tiny LNV}}$
are diagonalized simultaneously 
(and $h_{E}^{\mbox{\tiny LNV}}$ has two generation indices).
In particular, the smallness 
of relevant entries in $h_{D,\, E}^{\mbox{\tiny LNV}}$
is directly related to their smallness in the $h_{D,\, E}$ matrices.
The only new contributions 
to hadronic FCNC arise from loops, and in many cases (e.g., $B$
meson mixing) are suppressed as in the MSSM \cite{fb}.
(There are some differences 
due to the different masses of the Higgs and the leptonic
fields that propagate in the loops.)
In addition, the amplitudes for
leptonic FCNC processes, e.g.,
$\mu \rightarrow e\gamma$,
are typically proportional in our case to the mixing angle $\sin\zeta$ 
and/or to the electron mass \cite{hs}
(and, in general, depend sensitively on  $M_{SUSY}$).
Thus, they also lead to only mild constraints.
   
Charged current processes and constraints, e.g., 
weak current universality,
neutrinoless double-$\beta$ decay, etc., are induced at tree and loop
levels. They typically  constrain $h^{\mbox{\tiny LNV}} \lesssim 
{\cal{O}}(1 \times m_{f}/\nu)$, where $m_{f}$ is the relevant SM fermion mass
 (or lead to even weaker constraints).
Given the origin of $W^{\mbox{\tiny LNV}}$ in our case,
these constrains are typically satisfied and
could lead at most to an upper bound
on the ratio $\mu_{L}/\mu_{H} \sim \nu_{L}/\nu_{H} 
\sim h_{D,\, E}^{\mbox{\tiny LNV}}/h_{D,\, E}$.
(Note that not all  $h^{\mbox{\tiny LNV}}_{D_{ijk}}$ 
combinations are present in the basis
that  $h_{D}$ is  diagonal and that we do not specify 
any textures for the MSSM Yukawa matrices.)
Also, all constraints scale with the
inverse mass of the relevant (virtual) superpartner
that mediates the process, and are further weakened
as the mass parameters reach the few hundred GeV mark.
It is reasonable, however,  to require that the ratio 
$\mu_{L}/\mu_{H} \sim \nu_{L}/\nu_{H} \lesssim 1$
so that  the couplings in $W_{Y}^{\mbox{\tiny LNV}}$
are of the same order of magnitude (or smaller)
as the couplings in  $W_{Y}^{\mbox{\tiny MSSM}}$.

We conclude that a dedicated analysis and searches
could constrain the order parameters of the model:
$\sin\zeta$ (i.e., deviations from the alignment)
and the ratio $\mu_{L}/\mu_{H} \sim \nu_{L}/\nu_{H_{1}}$,
which  determine the gauge-lepton mixing and the size of the couplings in 
$W_{Y}^{\mbox{\tiny LNV}}$, respectively.
However, it will lead to only mild restrictions
on the model parameter space, and it is not called for
at present (given our poor knowledge of the supersymmetric 
spectrum parameters).
The former would then constrain the size of neutrino
masses and thus, serve in the future as a consistency
test of the model relating the neutrino and supersymmetric 
spectra.  
It could also constrain $\tan\beta$ (see above footnote).
The constraints on the ratio $\mu_{L}/\mu_{H}$ 
are, on the other hand, a function of $\tan\beta$ 
(recall that $\mu_{H}h_{D,\,E}^{\mbox{\tiny{LNV}}}/\mu_{L}
\approx  h_{D,\, E} = m_{d,\, e}/\nu\cos\beta$, 
where $m_{d,\, e}$ is the relevant fermion mass).

The ratio $\mu_{L}/\mu_{H}$
and the size of $h^{\mbox{\tiny{LNV}}}$
have also important implications for
collider signatures from exotic superpartner 
decays \cite{dawson,detal,rd,decays1,decays2},
in particular, from the decays of lightest superpartner (the LSP).
For example, there could be a single production 
of a scalar lepton in hadron
colliders which would then decay to two jets or two jets
and either like-sign dileptons or (degraded) missing energy 
(for example, see Ref.\ \cite{rd}).
Such decays serve, in general,  as a test of all 
supersymmetric lepton number violating
models (assuming that 
supersymmetry is established and characterized by experiment)
in both hadron and electron colliders.

Of particular interest is the case in which 
the scalar neutrino $\tilde{\nu}$ (that can  be singly produced)
is the LSP.
The gauginos and Higgsinos are then heavy
and the sneutrino  would decay exclusively to
two jets (assuming $h_{D} \gg h_{E}$).
Its charged $SU(2)$ partner, which has a similar mass,
will also decay preferably to jets. 
Superpartner decays
are characterized in that case by a $\geq 2j$ signal.
In the case of the Tevatron, for example,
the LSP dijet signal\footnote{
The elementary cross section
is given by ($d$ is any down-type quark and 
mass and family mixing  effects are neglected) \cite{detal}
\begin{center}
$\sigma(d_{i}\bar{d}_{i} \rightarrow \tilde{\nu} \rightarrow d_{j}\bar{d}_{j})
 = \frac{4\pi}{9} \frac{\Gamma_{i}\Gamma_{j}}{(\hat{s} - m_{\tilde{\nu}}^{2})^{2}
+m_{\tilde{\nu}}^{2}\Gamma_{total}^{2}},$
\end{center}
where $\hat{s}$ is the square of the parton center of mass energy
and, in our case,  
\begin{center}
$\Gamma_{i} = \frac{3}{16\pi}\left(\frac{\mu_{L}m_{d_{i}}}{\mu_{H}\nu\cos\beta}
\right)^{2}m_{\tilde{\nu}}.$
\end{center}
For simplictly, one  can assume  that the total width
$\Gamma_{total} \sim \Gamma_{j} \sim \Gamma_{b}$.}
could be only one order
of magnitude below the corresponding QCD signal \cite{detal}
(but would be further suppressed if the couplings are small).
However, the enhancement of the total dijet cross-section
is more significant (and is smeared over a larger energy range) 
if  there are more than one 
relevant threshold [e.g., its charged $SU(2)$ partner
and possibly scalars of all three lepton doublets].
These observations may require further attention\footnote{Similar
observations hold in the case of baryon number violation
via the operator $h^{\mbox{\tiny BNV}}UDD$ which is only weakly
constrained (in the absence of lepton number violation), 
but is forbidden here.} 
in view  of recent indications
of enhancements in the dijet inclusive cross section \cite{dijet}
and mass spectrum \cite{massspectrum} measurements.
In particular, the situation here
is quite different
than in $R$-parity  conserving supersymmetric theories,
where any new contributions to the inclusive dijet cross section are
only at the loop level and were shown recently
to be small or negative\cite{susydijet}.

\section{Astrophysics and Cosmology}
\label{sec:s6}

We concluded above that no significant constraints arise
on the model from laboratory and collider  data, but
stressed the unconventional decays and signatures of the supersymmetric 
particles. We also pointed out that the LSP is not expected to be stable.
Here, we will survey some astrophysical and cosmological aspects
of the model. 
We also elaborate on the LSP and neutrino decays.
In particular,  we show that the LSP (and thus, all other MSSM
superpartners)
is cosmologically irrelevant
in any supersymmetric model
with explicit (weak-scale) lepton number violation
which is the source of  the neutrino mass.
Radiative neutrino decays are shown to lead to only moderate constraints 
on the order parameter $\mu_{L}/\mu_{H}$ (unless $\tan\beta$ is large).
We do not consider here some 
issues which are more model dependent, e.g.,  that of the primordial
baryon asymmetry in the universe (but see, for example,  \cite{baryo,jn}).

$A$. {\it Neutrino oscillations:}
Current observations imply  not only massive neutrinos
but also neutrino mixing, so that the neutrinos could oscillate \cite{langrev}.
Indeed, we find  model dependent mixing
between the three neutrinos.
It was recently pointed out \cite{ralf,jn} that the mass and 
mixing patterns in the models 
with supersymmetric neutrino masses
are relevant for the solution of the solar 
and atmospheric neutrino problems \cite{langrev}.
Most importantly, we would like to stress that
the patterns that arise could be  quite different from those
in typical see-saw and similar models.
They are not proportional to any Yukawa couplings and the hierarchy
among the neutrino masses is not related to either the hierarchy
among the quarks or the charged leptons.

$B$.  {\it LSP decays:}
Since $R_{P}$ is effectively broken to the baryon parity in our models,
the LSP, i.e., the lightest neutralino or scalar, 
is not stable but decays \cite{hs,eg,dawson,bs,gg}
through small mixing (if it is a neutralino), 
and more probably,
via $h_{D}^{\mbox{\tiny LNV}}$  and 
$h_{E}^{\mbox{\tiny LNV}}$ interactions.
In order to render the LSP stable on a cosmological time scale
($\tau_{LSP} \sim 10^{17}$ sec)
one would have to fine tune  
\begin{equation}
h^{\mbox{\tiny LNV}} \sim h\frac{\mu_{L}}{\mu_{H}} \ll 
{\cal{O}}\left(\frac{10^{-6}}{\sqrt{\tau_{LSP} 
({\mbox{sec}})}}\right) \sim {\cal{O}}(10^{-14})
\label{lsp}
\end{equation}
for a neutralino LSP 
(and even more so for  a scalar LSP),  
and we neglected the decay via mixing which is further
suppressed. (See Ref.\ \cite{bs} for life time formulae).
It would imply, in our case, an unacceptable fine tuning
of $\mu_{L}/\mu_{H}$. But moreover, even if the neutrino masses
are generated by only Yukawa loops \cite{yukawaloop1,yukawaloop2},
they would be essentially massless 
given the constrain (\ref{lsp}), which is thus unacceptable.
We therefore conclude that the LSP decays on short time scales
and is cosmologically irrelevant.  (We find that a neutralino LSP  could, 
however, still be stable in our model on collider scales.)
Note that this argument cannot be cured by any symmetry that allows
for $W^{\mbox {\tiny LNV}} \neq 0$ since it is sufficient
to have a single lepton number 
violating Yukawa coupling that does not vanish.

$C$. {\it Neutrino mass and decays:}
The neutrino energy density and over-closure of the universe  
considerations constrain $m_{\nu_{\tau}} \lesssim 100$ eV \cite{100ev}
(unless it decays on cosmological time scales, 
i.e., $\tau_{\nu} \lesssim 1$ sec,
in which case 
one could also have $m_{\nu_{\tau}} \sim {\cal{O}}(10$ MeV)
\cite{roth}).
The neutrino decays, similarly to the LSP, via mixing to neutrinos
(i.e., $\nu \rightarrow 3\nu$) 
and via its small one-loop magnetic moment operator
$\propto (h^{\mbox{\tiny LNV}})^{2}$ (i.e., $\nu \rightarrow \nu\gamma$).
The former may be beneficial in the case of a ${\cal{O}}($MeV) neutrino
but is strongly suppressed. The latter is the dominant mode
with (adopting the calculation of Ref.\ \cite{yukawaloop2,emr} to our case)
\begin{equation}
\tau_{\nu} \sim 10^{7}\left(\frac{\mu_{H}}{\mu_{L}}\right)^{4}
\left(\frac{1 {\mbox{MeV}}}{m_{\nu}}\right)^{3}
\left(\frac{M_{SUSY}}{100 {\mbox{GeV}}}\right)^{6} {\mbox{ sec}},
\label{neutrino}
\end{equation}
where we assumed
$h^{\mbox{\tiny LNV}}  \sim h^{\mbox{\tiny LNV}}_{b} \sim 10^{-2}
\times(\mu_{L}/\mu_{H})$. 
[More generally, $10^{7}(\mu_{H}/\mu_{L})^{4}$ is replaced
by $10^{-1}(h^{\mbox{\tiny{LNV}}})^{-4}$.]
A comprehensive analysis of such decays was presented in Ref.\ \cite{emr}, 
leading to the constraint $\tau_{\nu}$(sec) $\gtrsim 10^{29}\, m_{\nu}$(MeV)
(for $m_{\nu} \lesssim 100$ eV).
From (\ref{neutrino}) one has the corresponding constraint
(for $m_{\nu} \lesssim 100$ eV),
\begin{equation}
\left(\frac{\mu_{H}}{\mu_{L}}\right)^{4}
\left(\frac{1 {\mbox{MeV}}}{m_{\nu}}\right)^{4}
\left(\frac{M_{SUSY}}{100 {\mbox{GeV}}}\right)^{6} \gtrsim 10^{22}.
\label{neutrino2}
\end{equation}
Again, no significant constraints arise
on the order parameter $\mu_{L}/\mu_{H}$.
(Note that for a given neutrino mass
only the combination $M_{SUSY}^{6}(\mu_{H}/\mu_{L})^{4}$
is constrained. For large $\tan\beta$, i.e., 
$h^{\mbox{\tiny{LNV}}} \rightarrow 1$, 
the constraints could be significant.)

$D$. {\it Dark matter sources:}
One might have hoped that the model would provide
a closed framework in which the primary candidates for cold
and hot dark matter, the LSP and the neutrino, respectively,
are related by mixing and/or decay chains and thus, have correlated
abundances. Such a scenario could make, for example,
critical density models for the universe more plausible.
However, we have concluded above 
that the LSP is cosmologically irrelevant
and the only (hot) dark matter 
candidates in the model  are the two heavier neutrinos
(which in some sense are the actual LSP). 
This is somewhat a disappointing aspect of the model.
Nevertheless, there could still be other sources for cold dark matter,
e.g., light components of the axion superfield 
(as was proposed in the case of a Peccei-Quinn  axion \cite{axiondm}).
Lastly, it is interesting to note that our models could reverse
the generic situation in the $\mu = {\cal{O}}(1$ GeV) 
lepton number conserving models which typically suffer from a slow
LSP annihilation rate \cite{fengetal}.

\section{Possible family dependences}
\label{sec:s7}

Throughout this work we have assumed universal $R$-charge assignment 
to all matter fields. As a result, the vector $\mu_{\alpha} =
(\mu_{H_{1}},\, \mu_{L}(1,\, 1,\, 1))$ was $SU(3)$ symmetric.
There are many ways in which the $SU(3)$ could be broken, most
obviously so, if the $R$-charge assignment is family dependent.

It was suggested that a family dependent assignment may be needed
in order to cancel the anomalies of a gauged $U(1)_{R}$ theory
and that it is related to the fermion mass problem
(but, as we noted above,  it could lead to dangerous $D$-terms).
In Ref.\ \cite{gaugeR} Yukawa operators involving the first two families
were forbidden by the assignment $R(\phi_{1}\phi_{2}\phi_{3}) > 2$
but the possibility of dynamical couplings was not discussed.
A more ambitious program was pursued in Ref.\ \cite{chan}
where it was suggested that all Yukawa couplings (but not $\mu_{\alpha}$)
are dynamical variables that depend on the $U(1)_{R}$
breaking parameter $\theta$ (in a similar manner to the 
horizontal symmetry approach \cite{hor}). 
However, being an $R$-symmetry complicates
the anomaly cancelation equations and Ref.\cite{chan} chose not to
consider (and satisfy) the complete set of equations, in particular,
those involving hidden fields.
Given the ambiguous status of the gauged $U(1)_{R}$,
we chose the simplest possible $U(1)_{R}$ charge assignments
and did not consider the MSSM Yukawa couplings as 
dynamical variables.

Nevertheless, 
it is very likely that the solution of the 
fermion mass problem, and in particular,
if it relies on some symmetry principles and/or if the Yukawa couplings
are dynamical variables, would break explicitly the $SU(3)$
symmetry in $W_{M}$ as well, e.g., by forbidding
certain field combinations.
It could also break the Higgs-lepton soft mass universality 
from a non-minimal Kahler functions 
or $D$-terms, in which case the dynamical
alignment would fail. However, the latter breaking is constrained (from FCNC)
to $SU(2) \times U(1)$ type breaking of the $SU(3)$, and one can impose
the additional condition that 
breaking  the above $SU(3)$ by the leptonic soft-terms
is negligible. For example,
this would be the case  if the $SU(3)$ is broken 
mainly in the right-handed lepton singlet sector
(but then $W_{M}$ is still $SU(3)$ symmetric),
or if it is a global or discrete symmetry
(and the 
Kahler function is, e.g., minimal).
Therefore, we  do not consider the $SU(3)$ symmetry as a necessary
result of our model.

The above discussion also affects the phenomenology of the model.
In particular, if 
$\mu_{L_{\mu,\, e}} \ll \mu_{L_{\tau}}$
and only $\tau$ number is (significantly) broken, then  the constraints
on $\mu_{L}/\mu_{H}$ are quite weak. (Models in that spirit were 
discussed, e.g., in  Ref.\ \cite{ma}.)

\section{Summary and Conclusions}
\label{sec:s8}

In conclusion, we have shown that the neutrino mass could arise
from a generalized supersymmetric mass term $\mu_{\alpha}L^{\alpha}H_{2}$
in the weak-scale superpotential, on the condition that Higgs-lepton 
universality in the scalar potential is broken only weakly. 
This is indeed the situation, for example, 
in universal models for the soft supersymmetry
breaking parameters.
 
The restricted form of $R$-parity breaking in the model 
was realized, as an example, 
in the framework of a spontaneously broken $U(1)_{R}$ symmetry,
which is often present in models of dynamical supersymmetry breaking.
The symmetry framework provides also a solution
to the generalized $\mu$-problem while suppressing
lepton number violation 
in the Yukawa interactions and in the Kahler potential.
In hidden sector models the $U(1)_{R}$ scale is an intermediate scale.
The neutrino mass is related to the intermediate scale physics
in that case
in a very different fashion than in see-saw models.
In particular, it could be a result of supersymmetry breaking.
It offers a new mechanism for the neutrino mass generation
in supergravity and superstring theories, which does not require
to introduce any intermediate-scale see-saw structure.

The neutrino mass suppression is achieved dynamically (and is sensitive
to $\tan\beta$). The resulting mass range for the neutrino agrees with
not only the weaker collider limits but also with cosmological considerations.
The dynamical suppression is typically (but not always) destroyed in
grand-unified models, models with an intermediate right-handed neutrino,
and in models with arbitrary lepton-number violating Yukawa couplings,
once renormalization effects are taken into account.
  
We were able to define the SM lepton number at the weak-scale 
(after all symmetries were broken). The resulting Yukawa superpotential has 
an accidental
$Z_{3}$ baryon-parity symmetry, but the Yukawa 
couplings are not arbitrary and, in general,
satisfy all experimental constraints. One can still 
constrain the models' two order parameters.
However, given the above and the large number of parameters
in supersymmetric models, the constraints
are mild  (unless $\tan\beta \gg 1$).

While considering the phenomenological implications
of the models, we noted that:
\begin{enumerate}  
\item
The patterns of neutrino masses that arise could be quite different
from those that arise in typical see-saw models.
\item
The LSP is not stable on cosmological time-scales, an observation
which holds more generally in supersymmetric models
with weak-scale origin of the  neutrino masses.
\item
If the LSP is a scalar neutrino it most probably decays
in the detector, enhancing jet production
and leading to a typical $\geq 2j$ signal.
\end{enumerate}  
These observations call for consideration of 
unorthodox scenarios when considering phenomenological
implications of supersymmetric models.

\acknowledgements

It is pleasure to thank Herbi Dreiner, Ralf Hempfling,
and Yossi Nir for discussions of their works,
and to  Paul Langacker and Masahiro Yamaguchi
for their comments on the manuscript.
This work was partially supported by a fellowship (N.\ P.) 
and grant No. SFB-375-95 (research in astroparticle physics)
of the Deutsche Forschungsgemeinschaft, and by European Union
grant No. SC1-CT92-0789 (H.-P.\ N.). 



\begin{thebibliography}{65}
\bibitem{langrev}
For reviews and references, see P. Langacker,
{\it Physics Beyond the Standard Model IV},
Proceedings of the  Conference,  
Lake Tahoe, Calif., 1994, eds.\
J. F. Gunion, T. Han, and J. Ohnemus 
(World Scientific, Singapore, 1995) p. 487,
hep-ph/9503327; 
J. W. F. Valle,  Invited talk given at the Fourth International
Workshop on Theoretical and Phenomenological 
Aspects of Underground Physics
(TAUP 95), Toledo, Spain, 1995,  FTUV-96-13, hep-ph/9602369. 
\bibitem{seesaw}
M. Gell-Mann, P. Ramond, and R. Slansky, in {\it Supergravity}, 
Proceedings of the Workshop, Stony Brook, New York, 1979,
eds.\ P. van Nieuwenhuizen and D. Freedman 
(North Holland, Amsterdam, 1979)
p. 315;
T. Yanagida, in 
{\it Proceedings of the Workshop on Unified Theories and Baryon Number
in the Universe}, Tsukuba, Japan, 1979, eds.\ A. Sawada nad A. Sugamoto
(KEK Reprot No. 79-18, Tsukuba, 1979).
\bibitem{nillesrev}
For review and references, see H.-P. Nilles, 
Phys. Rep. 110, 1  (1984); {\it Testing the Standard Model},
eds.\ M. Cvetic and P. Langacker (World Scientific, Singapore, 1991) p. 633;
{\it The Building Blocks of Creation},
eds.\ S. Raby and T. Walker (World Scientific, Singapore, 1994) p. 291.
\bibitem{hk}
H. E. Haber and G. L. Kane, Phys. Rep. 117 (1985) 75.
\bibitem{books}
P. Nath et al., {\it Applied N = 1 Supergravity} (World Scientific, Singapore, 1984);
G. G. Ross, {\it Grand Unified Theories} (Benjamin, New York, 1984);
R. N. Mohapatra, {\it Unifcation and Supersymmetry} 
(Springer, New York, 1986, 1992).
\bibitem{hs}
L. J. Hall and M. Suzuki, Nucl. Phys. B231 (1984) 419.
\bibitem{bn}
T. Banks, Y. Grossman, E. Nardi, and Y. Nir,
Phys. Rev. D 52 (1995) 5319.
After completion of this work we also received their paper
F. M. Borzumati, Y. Grossman, E. Nardi, and Y. Nir, 
WIS-96-21-PH, hep-ph/9606251.
\bibitem{lee}
I. Lee, Nucl. Phys. B246 (1984) 120.
\bibitem{aulakh} 
C. S. Aulakh and R. N. Mohapatra, Phys. Lett. 119B (1982) 136.
\bibitem{eg}
J. Ellis, G. Gelmini, C. Jarlskog, G. G. Ross, and J. W.  F. Valle,
Phys. Lett. 150B (1985) 142.
\bibitem{rv}
G. G. Ross and J. W.  F. Valle, Phys. Lett. 151B (1985) 375.
\bibitem{yukawaloop1}
K. S. Babu and R. N. Mohapatra, Phys. Rev. Lett. 64 (1990) 1705;
Phys. Rev. D 42 (1990) 3778; {\it ibid.} 43 (1991) 2278.
\bibitem{yukawaloop2}
R. Barbieri, M. M. Guzzo, A. Masiero, and D. Tommasini, 
Phys. Lett. B 252 (1990) 251; 
E. Roulet and D. Tommasini, {\it ibid.} 256 (1991) 218.
\bibitem{dawson}
S. Dawson, Nucl. Phys. B261 (1985) 297.
\bibitem{z2}
G. Farrar and P. Fayet, Phys. Lett. 76B (1978) 575.
\bibitem{z3}
L. E. Ibanez and G. G. Ross, Nucl. Phys. B368 (1992) 3.
\bibitem{kmn}
D. Kapetanakis, P. Mayr, and H.-P. Nilles, Phys. Lett. B 282 (1992) 95. 
\bibitem{muproblem}
J. E. Kim and H.-P. Nilles, Phys. Lett. 138B (1984) 150.
\bibitem{alex}
For a recent discussion, see, 
N. Polonsky and A. Pomarol, Phys. Rev. D 51 (1995) 6532.
\bibitem{gm}
G. F. Giudice and A. Masiero, Phys. Lett. B 206 (1988) 480.
\bibitem{nk}
E. J. Chun, J. E. Kim, and H.-P. Nilles, Nucl. Phys. B370 (1992) 105;
J. E. Kim and  H.-P. Nilles,  Mod. Phys. Lett. A 9 (1994) 3575.
\bibitem{muzero}
W. Buchmuller and D. Wyler, Phys. Lett. 121B (1983) 321;
L. J. Hall and L. Randall, Nucl. Phys. B352 (1991) 289;
M. Dine and D. A. McIntire, Phys. Rev. D 46 (1992) 2594.
\bibitem{pq}
R. Peccei and H. Quinn, Phys. Rev. Lett. 38 (1977) 1440.
\bibitem{cvlang}
M. Cvetic and P. Langacker,  IASSNS-HEP-95-90, hep-ph/9511378; 
IASSNS-HEP-95-113, hep-ph/9602424.
\bibitem{cm}
J. A. Casas and C. Munoz, Phys. Lett. B 306 (1993) 288.
\bibitem{dynamical}
I. Affleck, M. Dine, and N. Seiberg, Nucl. Phys. B256 (1985) 557.
\bibitem{ns}
A. E.  Nelson and N. Seiberg, Nucl. Phys. B416 (1994) 46.
\bibitem{fengetal}
J. L. Feng, N. Polonsky, and S. Thomas, Phys. Lett. B 370 (1996) 95.
\bibitem{axion}
S. Weinberg, Phys. Rev. Lett. 40 (1978) 223;
F. Wilczek, Phys. Rev. Lett. 40 (1978) 279.
\bibitem{kimrev}
For review and references, see J. E. Kim, Phys. Rep. 149 (1987) 1.
\bibitem{axionscale}
P. J. Steinhardt and M. S. Turner, Phys. Lett. 129B (1983) 51;
K. Babu, S. M. Barr, and D. Seckel, Phys. Lett. B 336 (1994) 213;
G. Dvali, IFUP-TH-21-95, hep-ph/9505253; 
M. Kawasaki, T. Moroi, and T. Yanagida, UT-730, hep-ph/9510461.
\bibitem{bpr}
J. Bagger, E. Poppitz, and L. Randall, Nucl. Phys. B426 (1994) 3. 
\bibitem{polonyi}
J. Polonyi,  Budapest Report No. KFKI-93 (1977). 
\bibitem{gaugeR}
A. H. Chamseddine and  H. Dreiner, Nucl. Phys. B458 (1996) 65;
D. J. Castano, D. Z. Freedman, and C. Manuel, Nucl. Phys. B461 (1996) 50. 
\bibitem{gravity}
S. Hawking, Comm. Math. Phys. 43 (1975) 199;
Phys. Lett. 195B (1987) 337;
G. V. Lavrelashvili, V. Rubakov, and P. Tinyakov, JETP Lett. 46 (1987) 167;
S. Giddings and A. Strominger, Nucl. Phys. B307 (1988) 854;
S. Coleman, {\it ibid.} B310 (1988) 643.
\bibitem{bd}
T. Banks and L. J. Dixon, Nucl. Phys. B307 (1988) 93.
\bibitem{pp}
G. D. Coughlan, W. Fischler, E. W. Kolb, S. Raby, and G. G. Ross,
\newline
Phys. Lett. 113B (1983) 59.
\bibitem{nkb}
T. Banks, D. B. Kaplan, and A. E. Nelson,
Phys. Rev. D 49 (1994) 779. 
\bibitem{louisk}
For general formulae for the soft supersymmetry breaking parameters 
and for references, see, for example,
V. S. Kaplunovsky and J. Louis, Phys. Lett. B 306 (1993) 269. 
\bibitem{halletal}
D. E. Brahm, L. J. Hall, and S. D. H. Hsu, 
Phys. Rev. D 42 (1990) 1860.
\bibitem{ralf}
R. Hempfling,  MPI-PhT/95-59, hep-ph/9511288.
\bibitem{aleph}
The ALEPH Collabortion, D. Buskulic et al., Phys. Lett. B 349 (1995) 585. 
\bibitem{100ev}
S. S. Gerstein and Ya. B. Zeldovich, Zh. Eksp. Teor. Fiz. Pis'ma 4 (1972) 174;
R. Cowsik and J. McClelland, Phys. Rev. Lett. 29 (1972) 669;
P. Hut, Phys. Lett. 96B (1977) 85;
K. Sato and H. Kobayashi, Prog. Theor. Phys. 58 (1977) 1775;
B. W. Lee and S. Weinberg, Phys. Rev. Lett. 39 (1977) 165;
M. I. Vysotsky, A. D. Doglov, and Ya. B. Zeldovich,  JETP Lett. 26 (1977) 188.
\bibitem{rges}
V. Barger, M. S. Berger, R. J. N. Phillips, and T. Wohrmann,
MADPH-95-910, hep-ph/9511473; 
Hempfling \cite{ralf};
H. Dreiner and H. Pois, ETH-TH/95-30, hep-ph/9511444;
B. de Carlos and P. L. White, SUSX-TH/96-003, hep-ph/9602381.
Note that the renormalization effects in the soft parameters
discussed by de Carlos and White do not appear in our case.
\bibitem{flat}
B. Gato, J. Leon, J. Perez-Mercader, and M. Quiros, Nucl. Phys. B260 (1985) 203.
\bibitem{bs}
A. Bouquet and P. Salati, Nucl. Phys. B284 (1987) 557, and refernces therein.
\bibitem{Majoron}
D. Comelli, A. Masiero, M. Pietroni, and A. Riotto, Phys. Lett. B 324 (1994) 397.
\bibitem{smir}
A. Yu Smirnov and F. Vissani, Phys. Lett. B 341 (1994) 173;
A. Brignole, H. Murayama, and R. Rattazzi, {\it ibid.} 335 (1994) 345.
\bibitem{constraints}
Hall and Suzuki \cite{hs};
Dawson \cite{dawson};
R. N. Mohapatra, Phys. Rev. D 34 (1986) 3457;
V. Barger, G. F. Giudice, and T. Han, Phys. Rev. D 40 (1989) 2987;
R. Barbieri, D. E. Brahm, L. J. Hall, and S. D. H. Hsu, Phys. Lett. B 238 (1990) 86;
Brahm et al. \cite{halletal}; 
I. Hinchliffe and T. Kaeding, Phys. Rev. D 47 (1993) 279;
L. Roszkowski, in {\it  Physics and  Experiments With Linear $e^{+}e^{-}$ Colliders},
eds.  F. A. Harris, S. L. Olsen, S. Pakvasa, and X. Tata
(World Scientific, Singapore, 1993) p. 854, hep-ph/9309208; 
G. Bhattachrayya, J. Ellis and K. Sridhar, Mod. Phys. Lett. A 10 (1995) 1583;
G. Bhattachrayya and D. Choudhury, {\it ibid.}  1699;
C. E. Carlson, P. Roy, and M. Sher, Phys. Lett. B 357 (1995) 99;
M. Hirsch, H. V. Klapdor-Kleingrothaus, and S. G. Kovalenko,
Phys. Rev. Lett. 75 (1995) 17; 
K. S. Babu and R. N. Mohapatra, {\it ibid.} 2276;
F. de Campos, M. A. Garcia-Jareno, A. S. Joshipura, J. Rosiek, 
and J. W. F. Valle, 
Nucl. Phys. B451 (1995) 3;
M. Nowakowski and  A. Pilaftsis, {\it ibid.} B461 (1996) 19;
Banks et al., \cite{bn};
K. Agashe and M. Graesser, LBL-37823, hep-ph/9510439; 
A. Yu. Smirnov and F. Vissani, IC-96-16, hep-ph/9601387; 
F. Vissani, to be published in 
{\it The proceedings of V Hellenic School and Workshops on Elementary Particle
Physics}, Corfu, Greece, Sept. 1995, IC-96-32, hep-ph/9602395; 
de Carlos and White \cite{rges};
D. Choudhury and P. Roy, MPI-PTh/96-20, hep-ph/9603363;
M. Chaichian and K. Huitu, HU-SEFT-R-1996-09, hep-ph/9603412.
\bibitem{fb} 
See, for example,
S. Bertolini, F. M. Borzumati, A. Masiero, and G. Ridolfi, 
Nucl. Phys. B353 (1991) 591.
\bibitem{detal}
S. Dimopoulos, R. Esmailzadeh, L. J. Hall, J.-P. Merlo, and G. D. Starkman,
Phys. Rev. D 41 (1990) 2099.
\bibitem{rd}
H. Dreiner and G. G. Ross, Nucl. Phys. B365 (1991) 597.
\bibitem{decays1}
H. Dreiner and R. J. N. Phillips, Nucl. Phys. B367 (1991) 591;
D. P. Roy, Phys. Lett. B 283 (1992) 270;
J. Butterworth and H. Dreiner, {\it ibid.} B397 (1993) 3; 
H. Dreiner, M. Guchait, and D. P. Roy, Phys. Rev. D 49 (1994) 3270;
V. Barger, M. S. Berger, P. Ohmann, and R. J. N. Phillips, Phys. Rev. D 50 (1994) 4299; 
G. Bhattachrayya, D. Choudhury, and K. Sridhar, 
Phys. Lett. B 355 (1995) 193;
M. Guchait and D. P. Roy, TIFR/TH/96-10, hep-ph/9603219.
\bibitem{decays2}
S. Dimopoulos and L. J. Hall, Phys. Lett. B 207 (1988) 210;
Barger et al. \cite{constraints};
R. Godbole, P. Roy, and X. Tata, Nucl. Phys. B401 (1993) 67;
V. Barger, W.-Y. Keung, and R. J. N. Phillips, Phys. Lett. B 364 (1995) 27; 
N. K. Krasnikov, hep-ph/9511464;
D. Choudhury CERN-TH/95-250, hep-ph/9511466.
\bibitem{dijet}
A. A. Bhatti, FERMILAB-CONF-95/192-E;
The CDF Collaboration, F. Abe et al.,
FERMILAB-PUB-96/020-E, hep-ex/9601008. 
\bibitem{massspectrum}
R. M. Harris, FERMILAB-CONF-95/152-E, hep-ex/9506008;
The CDF Collaboration, F. Abe et al., Phys. Rev. Lett. 74 (1995) 3538.  
\bibitem{susydijet}
V. Barger, M. S. Berger, and R. J. N. Phillips, 
MADPH-95-920, hep-ph/9512325;
P. Kraus and F. Wilczek, Caltech-68-2032, hep-ph/9601279;
J. Ellis and D. A. Ross, CERN-TH/96-108, hep-ph/9604432.
(The light gluino case was discussed by I. Terekhov and L. Clavelli,
UAHEP964, hep-ph/9603390.)
\bibitem{baryo}
B. A. Campbell, S. Davidson, J. Ellis, and K. A. Olive,
Phys. Lett. B 256 (1991) 457;
H. Dreiner and G. G. Ross, Nucl. Phys. B410 (1993) 188.
\bibitem{jn}
A. S. Joshipura and M. Nowakowski,  Phys. Rev. D 51 (1995) 2421;
{\it ibid.} 5271.
\bibitem{gg}
G. F. Giudice, 
in {\it Proceedings of The 
XII Warsaw Symposium on Elementary Particle Physics}, 
Kazimierz, Poland, 1989,
eds.\  Z. Ajduk, S. Pokorski, and A. K. Wroblewski 
(World Scientific, Singapore, 1990) p. 608.  
\bibitem{roth}
For review and references, see
I. Z. Rothstein, talk given at 
the Joint Meeting of the International Symposium on Particles, Strings and
Cosmology and 
the IXX  Johns Hopkins Workshop on Current Problems in Particle Theory,
Baltimore, MD, 1995,  hep-ph/9506443. 
\bibitem{emr}
K. Enqvist, A. Masiero, and A. Riotto, Nucl. Phys. B373 (1992) 95.
\bibitem{axiondm} 
K. Rajagopal, M. S. Turner, and F. Wilczek, Nucl. Phys. B358 (1991) 447;
J. E. Kim, Phys. Rev. Lett. 67 (1991) 3465;
E. J. Chun, J. E. Kim, and H.-P. Nilles, Phys. Lett. B 287 (1992) 123;
D. H. Lyth, Phys. Rev. D 48 (1993) 4523;
E. J. Chun and A. Lukas, Phys. Lett. B 357 (1995) 43;
S. Chang and H. B. Kim, UFIFT-HEP-96-8, hep-ph/9604222.
\bibitem{chan}
E. J. Chun, Phys. Lett. B 367 (1996) 226.
\bibitem{hor}
L. Ibanez and G. G. Ross, Phys. Lett. B 332 (1993) 100;
P. Binetruy and P. Ramond, {\it ibid.} 350 (1995) 49;
V. Jain and R. Shrock, {\it ibid.} 352 (1995) 83; ITP-SB-95-22, hep-ph/9507238;
Y. Nir, Phys. Lett. B 354 (1995) 107.
\bibitem{ma}
E. Ma and P. Roy, Phys. Rev. D 41 (1990) 988;
E. Ma, in {\it Proceedings of The
Tallinn Symposium on Neutrino Physics}, Lohusalu, Estonia, 1993,
eds. I. Ots and L. Palgi (Estonian Acad. Sci., Inst. Phys., 1994) p. 162,
hep-ph/9310280.


\end{thebibliography}
\end{document}